\documentclass[sigconf,nonacm]{acmart}

\renewcommand\footnotetextcopyrightpermission[1]{}

\usepackage{bm}
\usepackage{dsfont}

\usepackage{graphicx}
\usepackage{subcaption}

\usepackage{booktabs}
\usepackage{multirow}

\usepackage{algorithm}
\usepackage{algorithmic}

\usepackage{xcolor}
\usepackage{hyperref}

\newcommand{\FASCL}{\textsc{FASCL}}

\newcommand{\best}[1]{\textbf{#1}}
\newcommand{\second}[1]{\underline{#1}}

\begin{document}

%%
%% Title
%%
\title{Cross-Sectional Asset Retrieval via\texorpdfstring{\\}{ }Future-Aligned Soft Contrastive Learning}

%%
%% Authors
%%
\author{Hyeongmin Lee}
\authornote{Corresponding author.}
\affiliation{%
  \institution{Seoul National University of Science and Technology}
  \city{Seoul}
  \country{Republic of Korea}
}

\author{Chanyeol Choi}
\affiliation{%
  \institution{LinqAlpha}
  \city{Cambridge, MA}
  \country{United States}
}

\author{Jihoon Kwon}
\affiliation{%
  \institution{LinqAlpha}
  \city{Cambridge, MA}
  \country{United States}
}

\author{Yoon Kim}
\affiliation{%
  \institution{Massachusetts Institute of Technology}
  \city{Cambridge, MA}
  \country{United States}
}

\author{Alejandro Lopez-Lira}
\affiliation{%
  \institution{University of Florida}
  \city{Gainesville, FL}
  \country{United States}
}

\author{Wonbin Ahn}
\affiliation{%
  \institution{LG AI Research}
  \city{Seoul}
  \country{Republic of Korea}
}

\author{Justin Xu}
\affiliation{%
  \institution{MillTech}
  \city{London}
  \country{United Kingdom}
}

\author{Srijan Sood}
\affiliation{%
  \institution{J.P. Morgan AI Research}
  \city{New York, NY}
  \country{United States}
}

\author{Qingsong Wen}
\affiliation{%
  \institution{Squirrel Ai Learning}
  \city{Bellevue, WA}
  \country{United States}
}

\author{Chun-Li Yang}
\affiliation{%
  \institution{Cathay Securities Investment Trust Co., Ltd.}
  \city{Taipei}
  \country{Taiwan}
}

\author{Yongjae Lee}
\affiliation{%
  \institution{Ulsan National Institute of Science and Technology}
  \city{Ulsan}
  \country{Republic of Korea}
}

%% =============================================================================
%% ABSTRACT
%% =============================================================================
\begin{abstract}
Asset retrieval (finding similar assets in a financial universe) is central
to quantitative investment decision-making. Existing approaches define
similarity through historical price patterns or sector classifications, but
such backward-looking criteria provide no guarantee about future behavior. We
argue that effective asset retrieval should be \emph{future-aligned}: the
retrieved assets should be those most likely to exhibit correlated future
returns. To this end, we propose \textbf{Future-Aligned Soft Contrastive
Learning (\FASCL{})}, a representation learning framework whose soft
contrastive loss uses pairwise future return correlations as continuous
supervision targets. We further introduce an evaluation protocol designed to directly assess
whether retrieved assets share similar future trajectories. Experiments on
5,631 US-listed securities against 14 baselines show that \FASCL{}
attains the best future return correlation at every retrieval depth
and the best rank information coefficient at every depth and horizon,
leads on trend
consistency in 13 of 16 cells, and gives the highest gross Sharpe ratio in a
spread trading backtest at every basket size.  Code is available at
\url{https://github.com/HyeongminLEE/fascl}.
\end{abstract}

%%
%% CCS concepts (mandatory for ACM papers over two pages)
%%


%%
%% Keywords
%%

\maketitle

%% =============================================================================
%% 1. INTRODUCTION
%% =============================================================================
\section{Introduction}
\label{sec:introduction}

Financial machine learning has long been dominated by the forecasting
paradigm: directly predicting future prices or returns. Yet the
low signal-to-noise ratio~\cite{gu2020empirical} and non-stationarity of
financial markets limit the generalization of point-prediction
models, motivating a shift toward \emph{representation learning}: mapping
assets into embedding spaces that capture latent market structure rather than
predicting exact future values.

A natural application of such representations is \textbf{asset retrieval}:
given a query asset, finding other assets from a universe whose market
behavior is most similar. This task underpins portfolio construction, where
identifying behaviorally similar assets enables diversification beyond
static sector labels~\cite{lopez2020machine}; pairs trading, where
co-moving pairs form the basis of statistical
arbitrage~\cite{gatev2006pairs}; and risk management, where uncovering
hidden correlations reveals concentration risks invisible to sector-based
analysis~\cite{fama1993common}. Retrieval also offers inherent
explainability, presenting interpretable evidence:
\emph{``Asset A is expected to co-move with assets
B and C, so a position in A can be hedged or diversified with them.''}

But what does ``similar'' mean? Traditional methods rely on Pearson
correlation of historical returns, DTW-based pattern matching, or
sector/industry membership. These criteria are \emph{backward-looking}:
they identify assets that \emph{have} moved together, not those that
\emph{will}. We contend that, for asset retrieval to be practically useful,
the notion of similarity must be \textbf{future-aligned}: retrieved assets
should be those most likely to exhibit correlated future behavior.

Despite this intuitive appeal, existing \emph{retrieval-oriented}
methods neither \emph{train for} nor \emph{evaluate against} future
behavior similarity: self-supervised methods define positive pairs
through augmentation of the query itself~\cite{chen2020simclr,
yue2022ts2vec, woo2022cost, simstock2023} or temporal
neighborhood~\cite{tonekaboni2021tnc}---none grounded in future
outcomes---while evaluation relies on indirect downstream tasks such as
pairs-trading profitability~\cite{simstock2023} rather than directly
measuring future trajectory similarity.  Although recent prediction
studies have incorporated future outcome signals (e.g., binary price
direction~\cite{du2024explainable}), they target single-asset
forecasting rather than cross-sectional retrieval, and reduce future
behavior to coarse categorical labels.  The financial domain, however,
provides such a signal directly: future returns are continuous, observable
quantities that can rank assets by their degree of future co-movement.

We propose \textbf{Future-Aligned Soft Contrastive Learning (\FASCL{})},
a representation learning framework that closes both gaps. At its core is
a \emph{soft contrastive loss} in which pairwise future return
correlations (the Pearson correlation of daily return series over a future
horizon) serve as continuous supervision targets, transformed into a target
probability distribution via temperature-scaled softmax. The model minimizes
the KL divergence between this target distribution and the embedding
similarity distribution, so that pairs with highly correlated futures are
pulled together while uncorrelated pairs are pushed apart, all on a
continuous scale rather than through binary positive/negative assignments.

To evaluate whether retrieval is truly future-aligned, we introduce three
metrics that capture complementary aspects of future behavior consistency:
directional agreement (\emph{Trend Consistency@$K$}), trajectory-level
correlation (\emph{Future Return Correlation@$K$}), and cross-sectional
prediction quality derived from retrieval consensus (\emph{Information
Coefficient@$K$})~\cite{grinold2000active}. Together with Sector
Precision@$K$ for structural awareness, these form our
evaluation protocol for asset retrieval.

We conduct experiments on 5,631 US-listed securities spanning NASDAQ
and NYSE, comparing \FASCL{} against 14 baselines from five categories:
statistical methods, time series self-supervised models, forecasting-based
feature extractors, a financial domain model, and time series foundation
models (both zero-shot and fine-tuned). On the held-out test set,
\FASCL{} attains the best score in every FRC@$K$ and IC@$K$ cell and in
13 of 16 trend consistency cells; the three exceptions, at the 5-day
horizon with $K \geq 5$, trail historical Pearson correlation by at
most 0.6pp.

Our work makes three contributions toward establishing future-aligned asset
retrieval as a well-defined research problem:
\begin{enumerate}
    \item \textbf{Future-Aligned Soft Contrastive Loss.} We introduce a
    contrastive objective that uses pairwise future return correlations as
    continuous soft targets, aligning embedding similarity with future
    behavioral similarity so that retrieval reduces to nearest-neighbor
    search in the learned space.

    \item \textbf{Future Behavior Evaluation Metrics.} We propose TC@$K$,
    FRC@$K$, and IC@$K$---metrics that directly measure whether retrieved
    assets exhibit similar future behaviors across multiple horizons, moving
    beyond the indirect downstream evaluation prevalent in prior work.

    \item \textbf{Standardized Benchmark.} We design an
    evaluation framework with temporal data splits that prevent
    look-ahead bias, same-period retrieval constraints that reflect
    realistic investment scenarios, and 14 baselines from five method
    categories, all evaluated under an identical retrieval protocol.
    This provides, to our knowledge, the first comparison of these five
    families under a shared universe, temporal splits, and future-behavior
    evaluation.
\end{enumerate}

%% =============================================================================
%% 2. RELATED WORK
%% =============================================================================
\section{Related Work}
\label{sec:related}

\subsection{Financial Asset Similarity and Retrieval}

Traditional quantitative finance defines asset similarity through
backward-looking measures: Pearson correlation of historical
returns~\cite{fama1993common}, Dynamic Time Warping
(DTW)~\cite{sakoe1978dtw} for non-linear temporal alignment, or static
sector/industry classifications such as GICS (Global Industry
Classification Standard). These criteria capture whether
assets have co-moved in the past, but offer no guarantee of future
behavioral similarity.

Learning-based approaches have begun to address this limitation.
\textbf{SimStock}~\cite{simstock2023} pioneered systematic research on
stock similarity, employing a hypernetwork attention mechanism with triplet
loss to learn temporally invariant representations from rolling OHLCV (open, high,
low, close, volume) features. While it demonstrated that learned embeddings can identify
structurally similar stocks, its supervision is self-referential---the
triplet's positive and negative are both mixtures of the query window with
a dimension-shuffled copy of itself---and its evaluation is indirect
(pairs-trading profitability, index tracking error), which leaves open the
question of whether retrieved assets actually share similar \emph{future}
behaviors. Our work differs by using
future return correlations as explicit training supervision, evaluating
directly on future behavior consistency metrics, and operating on a
substantially larger universe (5,631 vs.\ ${\sim}$500 tickers).

A parallel line of work models inter-stock relationships through graph
neural networks (GNNs). HIST~\cite{xu2021hist} constructs predefined
and hidden concept graphs for trend forecasting, while temporal relational
methods~\cite{feng2019temporal} learn inter-stock dependencies for
prediction. However, these approaches are optimized for per-asset
forecasting rather than cross-sectional retrieval, and their reliance
on explicit graph construction scales quadratically with universe size,
making them less suited to efficient nearest-neighbor search over large
universes.
More recently, FinSeer~\cite{finseer2025} introduces a domain-specific
retriever for financial time series within a retrieval-augmented
generation (RAG) framework, but retrieval there serves as input
augmentation for language models rather than as the primary task.
THEME~\cite{theme2025} retrieves stocks for
a textual investment theme, so its queries are themes rather than assets.
Dolphin et al.~\cite{dolphin2024contrastive} learn asset embeddings
contrastively from financial time series, but take positives from
augmented views and evaluate on industry classification and portfolio
optimization rather than on future behavior.

Beyond retrieval, a growing body of work incorporates future outcome
signals into financial representation learning.
In industry classification, Bonne et al.~\cite{bonne2022peer} also
supervise with future co-movement, training models to predict next-year
return correlation from features of a \emph{pair} of companies.  A
pairwise scorer has no space to search: peers come from evaluating every
pair, which grows quadratically with the universe.  \FASCL{} places each
asset at a point instead, so the same embedding answers any query by
nearest-neighbor lookup.
Du et al.~\cite{du2024explainable} use contrastive learning for
explainable stock price movement prediction, defining positive pairs as
historical patterns that share the same \emph{binary} next-day price
direction and hard negatives as patterns with opposite future movement.
ContraSim~\cite{vinden2025contrasim} applies weighted contrastive
learning to financial news headlines, clustering days with similar
subsequent market responses.  Both demonstrate the value of
outcome-aligned supervision, yet they target single-asset
\emph{prediction} tasks (temporal pattern retrieval and market
forecasting, respectively) rather than \emph{cross-sectional} asset
retrieval.  Moreover, their supervision reduces future behavior to
coarse categorical labels (binary direction or market-level movement),
whereas \FASCL{} operates on \emph{continuous pairwise} future return
correlations derived solely from raw market data, preserving the graded,
ranking-based structure needed for retrieval at scale.

\subsection{Time Series Representation Learning}

Given the limitations of domain-specific financial methods, general-purpose
time series representation learning offers an alternative source of
embeddings for asset retrieval. We survey three families of approaches,
all of which we include as baselines.

\noindent\textbf{Contrastive methods.}
TS2Vec~\cite{yue2022ts2vec} performs hierarchical contrastive learning
with contextual consistency at both instance and temporal levels, using
augmented views of the same series as positives.
CoST~\cite{woo2022cost} extends this with disentangled seasonal-trend
representations through a momentum contrastive framework.
TNC~\cite{tonekaboni2021tnc} defines positives based on temporal
neighborhood (segments close in time). All three define positive pairs
through \emph{temporal proximity} or \emph{augmentation}, which captures
pattern self-similarity but carries no signal about future outcome
similarity. When repurposed for asset retrieval, their embeddings risk
grouping assets with visually similar patterns that subsequently diverge.

\noindent\textbf{Forecasting-based models.}
PatchTST~\cite{nie2023patchtst} applies a patch-based Transformer with
masked prediction or supervised forecasting.
TimesNet~\cite{wu2023timesnet} reshapes 1D series into 2D tensors via
FFT-detected periods to capture intra- and inter-period variations.
iTransformer~\cite{liu2024itransformer} inverts the standard Transformer
by applying attention across variates rather than time steps. When these
models' encoder representations are extracted for retrieval, they are
optimized for next-step prediction accuracy rather than inter-asset
behavioral similarity.

\noindent\textbf{Foundation models.}
MOMENT~\cite{goswami2024moment} is pre-trained on a large corpus of
public time series through masked reconstruction, while
Chronos~\cite{ansari2024chronos} tokenizes values into discrete bins and
trains a T5-based language model for probabilistic forecasting. Both
provide zero-shot embeddings, but their general-purpose objectives do not
specifically encode financial similarity structure and may not distinguish
assets that merely look alike from those that will co-move.

\subsection{Contrastive Learning Objectives}

The standard contrastive framework~\cite{oord2018cpc, chen2020simclr,
he2020moco} optimizes an InfoNCE-style loss with binary pair assignments:
each anchor has one (or a few) positives defined through data augmentation,
and all others are negatives. This binary treatment discards the
continuous structure of similarity among financial assets.
Supervised Contrastive Learning (SupCon)~\cite{khosla2020supcon} extends
this to discrete class labels, treating all same-class samples as
positives; yet applying SupCon to finance would require discretizing
returns into coarse categories.

Recent work has begun bridging this gap.
Rank-N-Contrast~\cite{zha2023rank} contrasts samples by their ranking in
the target space, preserving the ordinal structure of continuous labels, but
operates on scalar-valued targets for individual samples rather than
pairwise relational targets.
SoftCLT~\cite{lee2024softclt} generalizes binary contrastive assignments
to soft labels for time series, where the degree of ``positiveness''
is determined by \emph{input-space} proximity: temporal distance and
instance-level similarity (e.g., DTW distance).  While this preserves
the continuous nature of similarity, the supervision remains
\emph{observation-aligned}: it groups series that look alike in the past,
with no guarantee of similar \emph{future} behavior.

\FASCL{} adopts the same soft contrastive philosophy but redirects
the supervision source to \emph{pairwise future return
correlations}, a signal grounded in the outcome space rather than
the observation space.  This preserves the graded structure of
financial similarity and optimizes the embedding space for
future-aligned retrieval.  Our ablation study
(Section~\ref{sec:ablation}) confirms that the choice of supervision
source, not the soft formulation itself, is the key driver of
retrieval performance.

%% =============================================================================
%% 3. PROPOSED METHOD
%% =============================================================================
\section{Proposed Method}
\label{sec:method}

We present \FASCL{}, a framework for learning asset embeddings that
capture future behavioral similarity.
%% Figure~\ref{fig:architecture} illustrates the overall pipeline.
Given a universe of assets observed over a common time window, \FASCL{}
(i)~encodes each asset's raw market time series into a
fixed-dimensional embedding via a patch-based Transformer encoder, and
(ii)~trains the encoder with a soft contrastive loss that directly
aligns embedding similarity with pairwise future return correlation.
At inference time, retrieval reduces to nearest-neighbor search in
embedding space.

\subsection{Problem Formulation}
\label{sec:method:problem}

Consider a universe of $M$ assets $\mathcal{U} = \{a_1, \ldots, a_M\}$
observed at a common time point~$t$.  For each asset~$a_i$, we observe
a historical window $\mathbf{X}_i \in \mathbb{R}^{T \times C}$
spanning the preceding~$T$ trading days, where $C$~denotes the number
of market features.  Additionally, the subsequent~$H$ trading days
yield a daily return series
$\mathbf{r}_i = (r_{i,1}, \ldots, r_{i,H})$,
which serves as the supervision signal for future behavior.

\textbf{Asset retrieval}: given a query asset~$a_q$
with observation~$\mathbf{X}_q$, retrieve $K$~assets from the universe
whose future behavior is most similar to that of~$a_q$.
We learn an encoder
$f_\theta \colon \mathbb{R}^{T \times C} \to \mathbb{R}^{D}$
such that the similarity ranking of embeddings aligns with
the ranking of future return correlations:
\begin{equation}
  \label{eq:objective}
  \rho(\mathbf{r}_i, \mathbf{r}_j)
    > \rho(\mathbf{r}_i, \mathbf{r}_k)
  \;\Longrightarrow\;
  \mathrm{sim}\!\bigl(f_\theta(\mathbf{X}_i),\, f_\theta(\mathbf{X}_j)\bigr)
    > \mathrm{sim}\!\bigl(f_\theta(\mathbf{X}_i),\, f_\theta(\mathbf{X}_k)\bigr),
\end{equation}
where $\mathrm{sim}(\cdot,\cdot)$ denotes cosine similarity and
$\rho(\cdot,\cdot)$ the Pearson correlation of future daily returns.

\subsection{Encoder Architecture}
\label{sec:method:encoder}

We employ a Transformer encoder~\cite{vaswani2017attention} adapted for
multivariate financial time series, following the patch embedding
paradigm of ViT~\cite{dosovitskiy2021vit} and
PatchTST~\cite{nie2023patchtst}.

\noindent\textbf{Input normalization.}
Each window $\mathbf{X}_i \in \mathbb{R}^{T \times C}$ is z-score
normalized per feature across the time dimension:
$\tilde{x}_{t,c} = ({x}_{t,c} - \mu_c) / \sigma_c$,
where $\mu_c$ and $\sigma_c$ are the mean and standard deviation of
feature~$c$ within the window.  This removes absolute price and volume
scales, forcing the encoder to learn from relative patterns.

\noindent\textbf{Patch embedding.}
The normalized window
$\tilde{\mathbf{X}}_i \in \mathbb{R}^{T \times C}$
is split into $N = T/P$ non-overlapping patches of size~$P$ along the
time axis.  A shared linear projection (implemented as a 1D convolution
with kernel size and stride equal to~$P$) maps each
$P \times C$ patch to a $D$-dimensional token, followed by layer
normalization:
\begin{equation}
  \mathbf{e}_n = \mathrm{LN}\!\bigl(
    \mathrm{Conv1d}(\tilde{\mathbf{X}}_{\,(n-1)P+1\,:\,nP})
  \bigr),
  \quad n = 1, \ldots, N.
\end{equation}

\noindent\textbf{Positional encoding and [CLS] token.}
A learnable \texttt{[CLS]} token
$\mathbf{e}_0 \in \mathbb{R}^D$
is prepended to the patch sequence, and learnable positional embeddings
$\mathbf{P} \in \mathbb{R}^{(N+1) \times D}$ are added:
\begin{equation}
  \mathbf{H}^{(0)} =
  [\mathbf{e}_0;\; \mathbf{e}_1;\; \ldots;\; \mathbf{e}_N]
  + \mathbf{P}.
\end{equation}

\noindent\textbf{Transformer blocks.}
The token sequence is processed by $L$~pre-norm Transformer
blocks~\cite{vaswani2017attention}, each consisting of multi-head
self-attention (MHSA) and a position-wise feed-forward network (FFN)
with GELU activation:
\begin{align}
  \mathbf{H}^{(\ell')}   &= \mathbf{H}^{(\ell)}
     + \mathrm{MHSA}\!\bigl(\mathrm{LN}(\mathbf{H}^{(\ell)})\bigr),
  \label{eq:mhsa} \\
  \mathbf{H}^{(\ell+1)} &= \mathbf{H}^{(\ell')}
     + \mathrm{FFN}\!\bigl(\mathrm{LN}(\mathbf{H}^{(\ell')})\bigr),
  \label{eq:ffn}
\end{align}
for $\ell = 0, \ldots, L{-}1$, where LN denotes layer normalization
and FFN consists of two linear layers separated by GELU and dropout.

\noindent\textbf{Pooling.}
The final embedding is obtained by mean-pooling over the patch token
outputs (excluding \texttt{[CLS]}) after a final layer normalization:
\begin{equation}
  \label{eq:pooling}
  \mathbf{z}_i = \frac{1}{N}\sum_{n=1}^{N} \mathbf{h}_n^{(L)},
\end{equation}
where $\mathbf{h}_n^{(L)}$ denotes the $n$-th patch token of the
final Transformer block's normalized output.

In our experiments, we set $T{=}64$, $C{=}6$, $P{=}4$ ($N{=}16$
patches), $D{=}384$, $L{=}8$ blocks with 8 attention heads and FFN
expansion ratio~4, yielding approximately 14M parameters.

\subsection{Future-Aligned Soft Contrastive Loss}
\label{sec:method:scloss}

\noindent\textbf{Same-period batch construction.}
Each training batch is constructed by first sampling a random
date~$t$ from the training period, then sampling $B$~assets that have
sufficient data at date~$t$.  This ensures all samples within a batch
share the same temporal context, making pairwise future return
correlations a meaningful cross-sectional similarity signal.

\noindent\textbf{Target distribution.}
Given a batch of $B$ assets with future daily return series
$\{\mathbf{r}_1, \ldots, \mathbf{r}_B\}$
(each $\mathbf{r}_i \in \mathbb{R}^{H}$),
we compute the pairwise Pearson correlation matrix:
\begin{equation}
  \label{eq:pearson}
  C_{ij} = \frac{
    (\mathbf{r}_i - \bar{\mathbf{r}}_i)^\top
    (\mathbf{r}_j - \bar{\mathbf{r}}_j)
  }{
    \|\mathbf{r}_i - \bar{\mathbf{r}}_i\|\;
    \|\mathbf{r}_j - \bar{\mathbf{r}}_j\|
  }.
\end{equation}
For each anchor~$i$, a target probability distribution is constructed
via temperature-scaled softmax over the off-diagonal entries:
\begin{equation}
  \label{eq:target_dist}
  p_{ij} = \frac{
    \exp(C_{ij} / \tau_t)
  }{
    \sum_{k \neq i} \exp(C_{ik} / \tau_t)
  },
  \quad j \neq i,
\end{equation}
where $\tau_t$ is the \emph{target temperature} that controls the
sharpness of the distribution.

\noindent\textbf{Predicted distribution.}
The encoder embeddings
$\{\mathbf{z}_1, \ldots, \mathbf{z}_B\}$
induce a predicted distribution via cosine similarity:
\begin{equation}
  \label{eq:pred_dist}
  q_{ij} = \frac{
    \exp\!\bigl(\mathrm{sim}(\mathbf{z}_i, \mathbf{z}_j) / \tau\bigr)
  }{
    \sum_{k \neq i}
    \exp\!\bigl(\mathrm{sim}(\mathbf{z}_i, \mathbf{z}_k) / \tau\bigr)
  },
  \quad j \neq i,
\end{equation}
where $\tau$ is the \emph{loss temperature}.

\noindent\textbf{Loss.}
The soft contrastive loss minimizes the KL divergence between the
target and predicted distributions, averaged over all anchors:
\begin{equation}
  \label{eq:sc_loss}
  \mathcal{L}_{\mathrm{SC}}
  = \frac{1}{B}\sum_{i=1}^{B}
    D_{\mathrm{KL}}(\mathbf{p}_i \,\|\, \mathbf{q}_i)
  = \frac{1}{B}\sum_{i=1}^{B}\sum_{j \neq i}
    p_{ij} \log \frac{p_{ij}}{q_{ij}}.
\end{equation}

Unlike standard InfoNCE~\cite{oord2018cpc}, which assigns binary
positive/negative labels, $\mathcal{L}_{\mathrm{SC}}$ preserves
the continuous, graded nature of financial similarity: assets that
are 80\% correlated in future returns receive proportionally more
``pull'' than those at~30\%.  The separate temperatures~$\tau$
and~$\tau_t$ decouple the sharpness of the model's similarity
distribution from the target signal's scale.  We set $\tau{=}0.01$ and $\tau_t{=}0.05$.

\subsection{Training}
\label{sec:method:training}

The model is trained to minimize the soft contrastive loss
$\mathcal{L}_{\mathrm{SC}}$ (Eq.~\ref{eq:sc_loss}) using
AdamW~\cite{loshchilov2019decoupled}
($\beta_1 {=} 0.9$, $\beta_2 {=} 0.999$, weight decay~0.05) and
a cosine learning rate schedule with linear warmup over~10 epochs,
peak learning rate~$10^{-3}$, and minimum learning rate~$10^{-6}$.
Training runs for 100 epochs of 300 steps each, with batch
size~$B {=} 4{,}096$ and gradient clipping at norm~1.0.
Mixed-precision training (bfloat16) is used throughout.

\subsection{Retrieval}
\label{sec:method:retrieval}

At inference time, all assets in the evaluation universe are encoded
in a single forward pass.  Retrieval for a query asset~$a_q$ is
performed by computing cosine similarities between $\mathbf{z}_q$
and all corpus embeddings, then returning the top-$K$ most similar
assets subject to the constraints detailed in
Section~\ref{sec:evaluation}.

%% =============================================================================
%% 4. EVALUATION PROTOCOL
%% =============================================================================
\section{Evaluation Protocol}
\label{sec:evaluation}

\subsection{Dataset}
\label{sec:eval:dataset}

We construct our dataset from daily market data of 5,631 securities
listed on NASDAQ (3,443 tickers) and NYSE (2,188 tickers), collected
from public sources.  The universe is defined by exchange listing, so
alongside common stocks it contains a small number of exchange-traded
funds and other listed instruments, which we retain.  For each ticker, we
collect six features (open, high, low, close, volume, and trading
value) spanning 2010 to 2024.

To prevent information leakage, we adopt strict temporal splits:
\begin{itemize}
  \item \textbf{Train} (2010--2022): Same-period windows are sampled
  on-the-fly during training; 5,021 tickers pass the coverage filters.
  \item \textbf{Validation} (2023): Each ticker contributes the first
  and last 128 trading days of the year, each segment split into two
  consecutive 64-day windows.  Windows lacking 64 valid future days
  are dropped, yielding 20,728 evaluation samples.
  \item \textbf{Test} (2024): Same construction, yielding 16,370
  samples over 5,576 tickers.  The year's final window period lacks
  64 future days and is dropped entirely, so the samples span three
  window periods, of which the last two overlap by five trading days
  (our 2024 data span 251 trading days).  All main results are reported on this
  held-out set.
\end{itemize}
Because the soft contrastive loss requires $H{=}64$ future trading
days as supervision, training windows end in 2022 and their supervision
targets extend into the first quarter of 2023; both therefore precede
the test year by a wide margin.  The validation split, which we use
only for checkpoint selection, begins in January 2023 and so overlaps
the tail of that supervision.

\subsection{Retrieval Setup}
\label{sec:eval:setup}

We evaluate under \textbf{same-period retrieval}: each query retrieves
only from assets observed in the same time window.  This reflects
realistic investment scenarios in which all assets share the same
macro-economic context at the time of comparison.  We additionally
exclude the query's own ticker from retrieval candidates to avoid
trivial self-matches.

Similarity is measured by cosine similarity of encoder embeddings.
Top-$K$ retrieval ($K \in \{1, 5, 10, 20\}$) selects the $K$
highest-scoring candidates after applying the above constraints.

\begin{table*}[t]
  \centering
  \small
  \setlength{\tabcolsep}{3pt}
  \begin{tabular}{lcccccccc}
    \toprule
    & \multicolumn{4}{c}{FRC@$K$} & \multicolumn{4}{c}{SP@$K$ (\%)} \\
    \cmidrule(lr){2-5}\cmidrule(lr){6-9}
    Method
      & K=1 & K=5 & K=10 & K=20
      & K=1 & K=5 & K=10 & K=20 \\
    \midrule
%%GEN:tab_frc_sp BEGIN
    Random & 0.1408 & 0.1429 & 0.1431 & 0.1433 & 14.6 & 15.0 & 14.9 & 14.9 \\
    DTW & 0.2886 & 0.2624 & 0.2518 & 0.2418 & 27.8 & 25.5 & 24.4 & 23.4 \\
    Pearson Corr. & \second{0.3981} & \second{0.3737} & \second{0.3626} & \second{0.3506} & 42.8 & 40.5 & 38.6 & 36.7 \\
    \midrule
    TS2Vec & 0.2789 & 0.2570 & 0.2467 & 0.2366 & 28.7 & 26.2 & 25.2 & 24.2 \\
    CoST & 0.2827 & 0.2591 & 0.2513 & 0.2431 & 29.7 & 27.3 & 26.3 & 25.1 \\
    SoftCLT & 0.2913 & 0.2679 & 0.2575 & 0.2464 & 28.6 & 26.3 & 25.4 & 24.2 \\
    TNC & 0.1569 & 0.1579 & 0.1587 & 0.1577 & 17.4 & 16.9 & 17.0 & 16.9 \\
    \midrule
    PatchTST & 0.2525 & 0.2358 & 0.2278 & 0.2195 & 24.8 & 23.1 & 22.3 & 21.5 \\
    TimesNet & 0.2107 & 0.1980 & 0.1935 & 0.1890 & 21.2 & 20.8 & 20.4 & 20.0 \\
    iTransformer & 0.2477 & 0.2291 & 0.2207 & 0.2120 & 25.4 & 23.8 & 23.1 & 22.1 \\
    \midrule
    SimStock & 0.1871 & 0.1837 & 0.1819 & 0.1797 & \best{97.3} & \best{96.8} & \best{96.6} & \best{96.5} \\
    \midrule
    Chronos & 0.2684 & 0.2495 & 0.2405 & 0.2312 & 27.8 & 26.4 & 25.5 & 24.5 \\
    MOMENT & 0.2889 & 0.2660 & 0.2546 & 0.2438 & 29.1 & 27.3 & 26.0 & 24.8 \\
    MOMENT (FT) & 0.2948 & 0.2695 & 0.2575 & 0.2455 & 31.0 & 28.2 & 26.9 & 25.6 \\
    \midrule
    \FASCL{} (Ours) & \best{0.4427} & \best{0.4226} & \best{0.4083} & \best{0.3896} & \second{53.4} & \second{50.2} & \second{47.8} & \second{45.3} \\
    \midrule
    \emph{Oracle (upper bound)} & \emph{0.6905} & \emph{0.6489} & \emph{0.6244} & \emph{0.5963} & -- & -- & -- & -- \\
%%GEN:tab_frc_sp END
    \bottomrule
  \end{tabular}
  \caption{Test-set FRC@$K$ and SP@$K$.  \best{Bold} marks the best value
  in each column and \second{underline} the second-best, among methods.
  The oracle ranks candidates by their realized future return correlation
  under the same retrieval constraints, so it is an upper bound that no
  causal method can attain.}
  \vspace{-15pt}
  \label{tab:test_frc_sp}
\end{table*}

\begin{table*}[t]
  \centering
  \footnotesize
  \setlength{\tabcolsep}{3pt}
  \begin{tabular}{lcccccccccccccccc}
    \toprule
    & \multicolumn{4}{c}{TC@$K$ (1d)} & \multicolumn{4}{c}{TC@$K$ (5d)} &
      \multicolumn{4}{c}{TC@$K$ (20d)} & \multicolumn{4}{c}{TC@$K$ (60d)} \\
    \cmidrule(lr){2-5}\cmidrule(lr){6-9}\cmidrule(lr){10-13}\cmidrule(lr){14-17}
    Method
    & K=1 & K=5 & K=10 & K=20
    & K=1 & K=5 & K=10 & K=20
    & K=1 & K=5 & K=10 & K=20
    & K=1 & K=5 & K=10 & K=20 \\
    \midrule
%%GEN:tab_tc BEGIN
    Random & 48.8 & 48.2 & 48.2 & 48.3 & 58.4 & 58.6 & 58.7 & 58.9 & 52.9 & 53.3 & 53.2 & 53.2 & 51.6 & 52.1 & 52.0 & 52.0 \\
    DTW & 54.7 & 53.5 & 53.1 & 52.5 & 64.1 & 63.3 & 62.9 & 62.5 & 60.1 & 58.6 & 57.8 & 57.0 & 60.0 & 58.0 & 57.3 & 56.8 \\
    Pearson Corr. & \second{59.4} & \second{57.7} & \second{57.0} & \second{56.2} & \second{69.5} & \best{68.0} & \best{67.5} & \best{66.9} & \second{65.7} & \second{64.2} & \second{63.6} & \second{62.9} & \second{64.3} & \second{62.6} & \second{61.7} & \second{60.9} \\
    \midrule
    TS2Vec & 54.1 & 53.4 & 53.0 & 52.4 & 64.4 & 63.2 & 62.7 & 62.3 & 59.8 & 58.1 & 57.4 & 56.7 & 59.6 & 58.2 & 57.7 & 57.2 \\
    CoST & 54.4 & 53.7 & 53.1 & 52.6 & 64.3 & 63.3 & 63.0 & 62.6 & 59.3 & 58.2 & 57.7 & 57.3 & 59.4 & 58.0 & 57.5 & 57.0 \\
    SoftCLT & 55.3 & 53.8 & 53.1 & 52.5 & 64.3 & 63.7 & 63.3 & 62.8 & 59.5 & 58.3 & 57.8 & 57.2 & 60.2 & 58.3 & 57.7 & 57.0 \\
    TNC & 48.8 & 48.8 & 48.8 & 48.9 & 60.2 & 59.4 & 59.2 & 59.2 & 53.7 & 53.5 & 53.3 & 53.3 & 53.0 & 53.0 & 52.8 & 52.9 \\
    \midrule
    PatchTST & 53.0 & 52.1 & 51.6 & 51.2 & 62.5 & 62.0 & 61.7 & 61.3 & 59.1 & 57.9 & 57.4 & 56.8 & 58.1 & 57.0 & 56.6 & 56.1 \\
    TimesNet & 51.7 & 51.3 & 50.9 & 50.7 & 61.4 & 61.3 & 61.0 & 60.7 & 56.2 & 55.4 & 55.0 & 54.7 & 55.4 & 54.6 & 54.3 & 54.2 \\
    iTransformer & 53.2 & 52.0 & 51.8 & 51.4 & 63.2 & 62.5 & 62.2 & 61.7 & 57.7 & 56.9 & 56.5 & 56.0 & 57.4 & 56.7 & 56.0 & 55.5 \\
    \midrule
    SimStock & 49.7 & 49.2 & 49.2 & 49.1 & 60.1 & 60.1 & 60.2 & 60.2 & 55.1 & 54.6 & 54.7 & 54.6 & 55.7 & 55.1 & 55.0 & 54.8 \\
    \midrule
    Chronos & 54.2 & 53.4 & 52.9 & 52.6 & 63.6 & 63.0 & 62.6 & 62.2 & 58.9 & 57.7 & 57.2 & 56.6 & 58.7 & 57.6 & 57.4 & 56.9 \\
    MOMENT & 55.3 & 53.6 & 53.0 & 52.4 & 65.0 & 63.4 & 63.1 & 62.7 & 59.6 & 58.4 & 57.6 & 57.1 & 60.0 & 58.7 & 58.0 & 57.3 \\
    MOMENT (FT) & 55.3 & 53.8 & 53.3 & 52.6 & 64.1 & 63.6 & 62.9 & 62.5 & 59.7 & 58.7 & 57.9 & 57.2 & 60.2 & 58.5 & 57.8 & 57.2 \\
    \midrule
    \FASCL{} (Ours) & \best{60.4} & \best{58.8} & \best{58.1} & \best{57.2} & \best{69.7} & \second{67.6} & \second{66.9} & \second{66.5} & \best{67.1} & \best{64.6} & \best{64.0} & \best{63.5} & \best{66.0} & \best{63.5} & \best{62.6} & \best{62.0} \\
%%GEN:tab_tc END
    \bottomrule
  \end{tabular}
  \caption{Test-set TC@$K$ (\%) across future horizons of 1, 5, 20, and 60 trading days.
  The Random row reflects market drift, not a 50\% floor.}
  \vspace{-15pt}
  \label{tab:test_tc}
\end{table*}

\subsection{Metrics}
\label{sec:eval:metrics}

We evaluate retrieval quality through four complementary metrics.

\noindent\textbf{Trend Consistency@$K$ (TC@$K$).}
The fraction of top-$K$ retrieved assets whose future cumulative
return has the same sign (direction) as the query's:
\begin{equation}
  \label{eq:tck}
  \mathrm{TC@}K = \frac{1}{|\mathcal{Q}|}
  \sum_{q \in \mathcal{Q}} \frac{1}{K}
  \sum_{k=1}^{K}
  \mathds{1}\!\bigl[\mathrm{sign}(R_q^{(h)})
                   = \mathrm{sign}(R_k^{(h)})\bigr],
\end{equation}
where $R^{(h)}$ denotes the cumulative return over horizon~$h$.
We report TC@$K$ for $h \in \{1, 5, 20, 60\}$ days.
A drift-free random baseline would sit at 50\%; empirically, random
retrieval ranges from 48\% to 59\% across horizons
(Table~\ref{tab:test_tc}), as market-wide drift shifts the
probability of sign agreement.

\noindent\textbf{Future Return Correlation@$K$ (FRC@$K$).}
The average Pearson correlation between the query's and retrieved
assets' future daily return series, over all query--retrieval pairs at
depth~$K$:
\begin{equation}
  \label{eq:frck}
  \mathrm{FRC@}K = \frac{1}{|\mathcal{P}_K|}
  \sum_{(q,k) \in \mathcal{P}_K}
  \mathrm{Pearson}(\mathbf{r}_q,\; \mathbf{r}_k),
\end{equation}
where $\mathbf{r}_q, \mathbf{r}_k \in \mathbb{R}^{H}$ are the
$H$-day future daily return series and $\mathcal{P}_K$ is the set of
pairs $(q, k)$ with $q \in \mathcal{Q}$ and $k$ among the top-$K$
retrievals for~$q$, excluding pairs with an incomplete future series.
FRC@$K$ captures trajectory-level similarity beyond directional
agreement.  A random baseline achieves ${\approx}0.14$ rather than
exactly zero, reflecting the market-wide common factor (market beta)
that induces positive correlation among equities in general.

\noindent\textbf{Information Coefficient@$K$ (IC@$K$).}
For each query, the equal-weight mean of its top-$K$ peers' future
returns serves as a \emph{consensus prediction}:
\begin{equation}
  \hat{y}_q^{(h)} = \frac{1}{K}\sum_{k=1}^{K} R_k^{(h)}.
\end{equation}
IC@$K$ is the Spearman rank correlation between these consensus
predictions and the actual query returns, computed cross-sectionally
over all queries:
\begin{equation}
  \label{eq:ick}
  \mathrm{IC@}K = \mathrm{Spearman}\!\Bigl(
    \bigl\{\hat{y}_q^{(h)}\bigr\}_{q \in \mathcal{Q}},\;
    \bigl\{R_q^{(h)}\bigr\}_{q \in \mathcal{Q}}
  \Bigr).
\end{equation}
Unlike TC@$K$ and FRC@$K$, which assess per-pair quality, IC@$K$
measures whether retrieval-derived consensus separates assets
cross-sectionally.  The consensus is formed from the peers' realized
returns, so IC@$K$ diagnoses how well retrieval groups assets that move
together; it is not a forecast and is not tradable as stated.
We report IC@$K$ for $h \in \{1, 5, 20, 60\}$ days; because IC@$K$ uses the
Spearman correlation, IC@$K$ is a rank information coefficient.

\noindent\textbf{Sector Precision@$K$ (SP@$K$).}
The fraction of top-$K$ retrieved assets belonging to the same
sector as the query, using the exchange-provided sector taxonomy from
the NASDAQ stock screener.  This auxiliary metric assesses structural
awareness: behaviorally similar assets may reside in different sectors
and vice versa.  Sector labels cover
86\% of test samples (the remainder are delisted tickers absent from
the screener), and SP@$K$ is computed over the labeled subset.

\begin{table*}[t]
  \centering
  \footnotesize
  \setlength{\tabcolsep}{3pt}
  \begin{tabular}{lcccccccccccccccc}
    \toprule
    & \multicolumn{4}{c}{IC@$K$ (1d)} & \multicolumn{4}{c}{IC@$K$ (5d)} &
      \multicolumn{4}{c}{IC@$K$ (20d)} & \multicolumn{4}{c}{IC@$K$ (60d)} \\
    \cmidrule(lr){2-5}\cmidrule(lr){6-9}\cmidrule(lr){10-13}\cmidrule(lr){14-17}
    Method
    & K=1 & K=5 & K=10 & K=20
    & K=1 & K=5 & K=10 & K=20
    & K=1 & K=5 & K=10 & K=20
    & K=1 & K=5 & K=10 & K=20 \\
    \midrule
%%GEN:tab_ic BEGIN
    Random & 0.0611 & 0.0730 & 0.1002 & 0.1253 & 0.2104 & 0.3055 & 0.3430 & 0.3712 & 0.0291 & 0.0767 & 0.1050 & 0.1265 & 0.0415 & 0.0746 & 0.0836 & 0.0952 \\
    DTW & 0.1415 & 0.1792 & 0.1955 & 0.2050 & 0.3181 & 0.3937 & 0.4172 & 0.4314 & 0.1783 & 0.2142 & 0.2290 & 0.2328 & 0.1703 & 0.1696 & 0.1736 & 0.2009 \\
    Pearson Corr. & \second{0.2490} & \second{0.2756} & \second{0.2908} & \second{0.2999} & \second{0.4180} & \second{0.4745} & \second{0.4902} & \second{0.4972} & \second{0.2934} & \second{0.3448} & \second{0.3649} & \second{0.3870} & \second{0.2596} & \second{0.3010} & \second{0.3075} & \second{0.3198} \\
    \midrule
    TS2Vec & 0.1533 & 0.1991 & 0.2193 & 0.2281 & 0.3266 & 0.4065 & 0.4239 & 0.4405 & 0.1732 & 0.2121 & 0.2321 & 0.2431 & 0.1651 & 0.1931 & 0.2129 & 0.2294 \\
    CoST & 0.1497 & 0.2031 & 0.2279 & 0.2372 & 0.3242 & 0.4049 & 0.4324 & 0.4504 & 0.1666 & 0.2129 & 0.2476 & 0.2653 & 0.1742 & 0.1881 & 0.2072 & 0.2209 \\
    SoftCLT & 0.1645 & 0.1989 & 0.2050 & 0.2190 & 0.3248 & 0.4060 & 0.4332 & 0.4434 & 0.1598 & 0.2128 & 0.2373 & 0.2434 & 0.1760 & 0.1893 & 0.2013 & 0.2163 \\
    TNC & 0.0607 & 0.0859 & 0.1119 & 0.1493 & 0.2530 & 0.3435 & 0.3781 & 0.4002 & 0.0607 & 0.0870 & 0.0983 & 0.1289 & 0.0620 & 0.0803 & 0.0971 & 0.1180 \\
    \midrule
    PatchTST & 0.1166 & 0.1471 & 0.1605 & 0.1729 & 0.2872 & 0.3603 & 0.3936 & 0.4125 & 0.1474 & 0.1896 & 0.2111 & 0.2290 & 0.1425 & 0.1659 & 0.1798 & 0.1932 \\
    TimesNet & 0.1013 & 0.1572 & 0.1715 & 0.1863 & 0.2656 & 0.3621 & 0.3918 & 0.4065 & 0.1033 & 0.1445 & 0.1663 & 0.1870 & 0.0943 & 0.1170 & 0.1304 & 0.1439 \\
    iTransformer & 0.1310 & 0.1737 & 0.1985 & 0.2101 & 0.3146 & 0.3938 & 0.4161 & 0.4289 & 0.1406 & 0.1878 & 0.2113 & 0.2306 & 0.1304 & 0.1584 & 0.1754 & 0.1885 \\
    \midrule
    SimStock & 0.0777 & 0.1051 & 0.1360 & 0.1573 & 0.2480 & 0.3421 & 0.3850 & 0.4116 & 0.0910 & 0.1303 & 0.1632 & 0.1842 & 0.1088 & 0.1379 & 0.1534 & 0.1799 \\
    \midrule
    Chronos & 0.1408 & 0.1858 & 0.2046 & 0.2230 & 0.3152 & 0.3945 & 0.4186 & 0.4405 & 0.1621 & 0.2141 & 0.2351 & 0.2437 & 0.1514 & 0.1942 & 0.2131 & 0.2313 \\
    MOMENT & 0.1562 & 0.1934 & 0.2156 & 0.2279 & 0.3342 & 0.4082 & 0.4316 & 0.4427 & 0.1576 & 0.2089 & 0.2223 & 0.2437 & 0.1729 & 0.2018 & 0.2186 & 0.2338 \\
    MOMENT (FT) & 0.1607 & 0.1993 & 0.2194 & 0.2330 & 0.3318 & 0.4060 & 0.4253 & 0.4415 & 0.1764 & 0.2311 & 0.2404 & 0.2472 & 0.1883 & 0.2059 & 0.2111 & 0.2172 \\
    \midrule
    \FASCL{} (Ours) & \best{0.2910} & \best{0.3400} & \best{0.3423} & \best{0.3375} & \best{0.4586} & \best{0.4985} & \best{0.5056} & \best{0.5126} & \best{0.3548} & \best{0.3949} & \best{0.4067} & \best{0.4220} & \best{0.3162} & \best{0.3548} & \best{0.3660} & \best{0.3698} \\
%%GEN:tab_ic END
    \bottomrule
  \end{tabular}
  \caption{Test-set IC@$K$ across future horizons of 1, 5, 20, and 60 trading days.
  Higher values indicate greater cross-sectional ranking quality.}
  \vspace{-15pt}
  \label{tab:test_ic}
\end{table*}

\subsection{Baselines}
\label{sec:eval:baselines}

We compare against 14 baselines spanning five categories, all
evaluated under an identical same-period retrieval protocol:

\noindent\textbf{Statistical methods} (train-free).
\emph{Random} assigns random embeddings as a lower bound.
\emph{Pearson} uses historical close-price change vectors, so that
retrieval reduces to Pearson correlation of observed price movements.
\emph{DTW}~\cite{sakoe1978dtw} ranks candidates directly by Dynamic
Time Warping distance between normalized close price series, without
an embedding step.

\noindent\textbf{Time series self-supervised learning.}
\emph{TS2Vec}~\cite{yue2022ts2vec},
\emph{CoST}~\cite{woo2022cost}, and
\emph{TNC}~\cite{tonekaboni2021tnc}---contrastive methods that
define positives through temporal proximity or augmentation---are
trained on the same training split using their original objectives.
\emph{SoftCLT}~\cite{lee2024softclt} extends TS2Vec with soft
contrastive weights: instance-wise weights derived from pairwise
distances between input windows and temporal weights from time lags,
following the official implementation, with the same encoder
architecture and training protocol as TS2Vec.

\noindent\textbf{Forecasting-based models.}
\emph{PatchTST}~\cite{nie2023patchtst},
\emph{TimesNet}~\cite{wu2023timesnet}, and
\emph{iTransformer}~\cite{liu2024itransformer} are trained for
multi-step forecasting on the training split; encoder representations
are extracted for retrieval.

\noindent\textbf{Financial domain model.}
\emph{SimStock}~\cite{simstock2023} is trained with a triplet loss whose
positive and negative are both formed from the query window by mixing it
with a dimension-shuffled copy of itself ($\lambda{=}0.7$), and, following
its original design, additionally takes each asset's sector as a
categorical input embedding.

\noindent\textbf{Foundation models.}
\emph{MOMENT}~\cite{goswami2024moment} and
\emph{Chronos}~\cite{ansari2024chronos} provide pre-trained
embeddings used zero-shot.  \emph{MOMENT (FT)} additionally
fine-tunes all 342M parameters of MOMENT on our training split with
its forecasting objective, isolating whether fine-tuning on
in-domain data closes the gap.

All training-based baselines are trained on the same temporal split
(2010--2022) and evaluated under identical same-period retrieval
constraints.  For every objective that regresses raw values (the
forecasting models and the MOMENT fine-tune), inputs and targets are
normalized per window with the input window's channel statistics and
winsorized at $10\sigma$; without this, the volume and trading-value channels
dominate the MSE by several orders of magnitude.

%% =============================================================================
%% 5. EXPERIMENTS
%% =============================================================================
\section{Experiments}
\label{sec:experiments}

\subsection{Experimental Setup}
All reported results are on the \textbf{test split} (2024) under the same-period
retrieval protocol described in Section~\ref{sec:evaluation}.  Our final model
is \textbf{\FASCL{}} trained with the \emph{soft contrastive loss},
using input windows of length $T{=}64$ and $H{=}64$ future
days.  Training runs for 100 epochs with batch size 4,096, AdamW optimizer,
and cosine learning rate schedule (warmup 10 epochs).  All baselines are
trained and evaluated under identical data splits and retrieval constraints.

\begin{table}[t]
  \centering
  \footnotesize
  \setlength{\tabcolsep}{4pt}
  \begin{tabular}{lcccc}
    \toprule
    Variant & K=1 & K=5 & K=10 & K=20 \\
    \midrule
%%GEN:tab_ablation BEGIN
    \FASCL{} & 0.4427 & 0.4226 & 0.4083 & 0.3896 \\
    \;+ auxiliary regression head & 0.4435 & 0.4200 & 0.4067 & 0.3910 \\
    \midrule
    Observation-aligned targets & 0.2875 & 0.2642 & 0.2551 & 0.2467 \\
    Hard InfoNCE targets & 0.4259 & 0.3993 & 0.3855 & 0.3697 \\
    Return regression only & 0.1667 & 0.1601 & 0.1575 & 0.1548 \\
    Loss temperature $\tau{=}0.07$ & 0.4398 & 0.4185 & 0.4050 & 0.3885 \\
    \midrule
    CLS pooling & 0.4496 & 0.4379 & 0.4194 & 0.3966 \\
    With projection head & 0.3794 & 0.3517 & 0.3385 & 0.3248 \\
    Close price only & 0.4226 & 0.3994 & 0.3865 & 0.3722 \\
%%GEN:tab_ablation END
    \bottomrule
  \end{tabular}
  \caption{Ablation on test-set FRC@$K$.  \FASCL{} uses the soft
  contrastive loss alone; the second row adds an auxiliary
  return-regression head.  Every row below the first rule keeps that
  combined configuration and modifies exactly one component:
  supervision and loss (middle block) or architecture and input
  (bottom block).}
  \vspace{-10pt}
  \label{tab:ablation}
\end{table}

\subsection{Main Retrieval Results}
\label{sec:main_results}
Tables~\ref{tab:test_frc_sp}--\ref{tab:test_ic} report test-set performance
across all evaluation metrics on the held-out \textbf{test split} (2024).
\best{Bold} denotes the best value in each column and \second{underline}
the second-best.

\noindent\textbf{FRC@$K$ (Table~\ref{tab:test_frc_sp}).}
\FASCL{} achieves the highest FRC@$K$ at every~$K$, reaching 0.443 at
$K{=}1$ (+11\% relative over the second-best Pearson correlation at
0.398).  The margin is statistically significant at every~$K$:
a paired bootstrap over queries ($N{=}1{,}000$ resamples) gives
non-overlapping 95\% confidence intervals (at $K{=}1$,
0.443 [0.438, 0.448] vs.\ 0.398 [0.393, 0.404]) with $p < 0.001$.
Historical return correlation
already partially predicts near-future co-movement through momentum
effects~\cite{jegadeesh1993momentum}; explicit future-aligned
optimization improves on this strong baseline.
The strongest methods whose notion of similarity is grounded in the
observation space land in a narrow band well below Pearson: zero-shot
foundation models (MOMENT 0.289, Chronos 0.268), the self-supervised methods
(SoftCLT 0.291, CoST 0.283, TS2Vec 0.279), and even training-free DTW
(0.289) all sit between 0.26 and 0.30 at $K{=}1$.  The ceiling they
share indicates
that how alike two windows \emph{look} carries only limited
information about how their assets will co-move.  Fine-tuning all
342M parameters of MOMENT on our training split lifts its FRC@$K$ by only 2\%
(0.295), so in-domain adaptation under a forecasting objective does
not escape this ceiling either.
Representations extracted from forecasting models trail even this
band (PatchTST 0.253, iTransformer 0.248, TimesNet 0.211), and
SimStock yields 0.187: its triplet is built from corrupted copies of the
query alone, so nothing in its objective relates one asset to another.
To place these values on an absolute scale, the last row reports an
oracle that ranks candidates by their realized future return
correlation.  Its FRC@1 of 0.691 is the ceiling any method could reach
here, and \FASCL{} closes 54--55\% of the distance from random retrieval to
that ceiling at every~$K$, against 46--47\% for Pearson correlation.

\noindent\textbf{SP@$K$ (Table~\ref{tab:test_frc_sp}).}
SimStock achieves near-ceiling sector precision (97.3\% at $K{=}1$),
which is largely by construction since sector is one of its inputs,
yet it sits near the bottom on FRC@$K$ at every~$K$: encoding sector
membership almost perfectly does not make retrieval
future-behavior-consistent.
Conversely, \FASCL{} achieves the highest SP@$K$ among all methods
that receive no sector information (53.4\% at $K{=}1$ vs.\ 42.8\%
for Pearson), suggesting that same-sector assets share correlated
futures through common factor exposures, which \FASCL{} captures
without labels (see Section~\ref{sec:visualization}).

\noindent\textbf{TC@$K$ (Table~\ref{tab:test_tc}).}
\FASCL{} achieves the highest TC@$K$ at the 1-, 20-, and 60-day
horizons at every~$K$ (e.g., 67.1\% vs.\ 65.7\% for Pearson at
$K{=}1$, 20-day), and at the 5-day horizon at $K{=}1$.
At the 5-day horizon with $K \geq 5$, Pearson edges \FASCL{} by
0.4--0.6pp, consistent with short-horizon sign agreement being
dominated by momentum, which historical correlation measures
directly.  \FASCL{}'s advantage concentrates where it matters for
portfolio construction: at the 20- and 60-day horizons every
baseline trails it at every~$K$.

\noindent\textbf{IC@$K$ (Table~\ref{tab:test_ic}).}
\FASCL{} achieves the highest IC@$K$ at all horizons and~$K$ values,
with relative margins over Pearson from +3\% (5-day, $K{\geq}10$) to
+23\% (1-day, $K{=}5$).  The margins are narrowest at the 5-day
horizon and widen toward both shorter and longer horizons (at 20-day
with $K{=}5$, 0.395 vs.\ 0.345).  These gains indicate
that \FASCL{}'s retrieval neighborhoods yield a cross-sectionally
coherent ranking signal that holds across investment horizons
without horizon-specific tuning.

\noindent\textbf{Binary-relevance view.}
FRC@$K$ and IC@$K$ score retrieval on a continuous scale.  Casting the
same task in standard IR terms, we call a retrieved asset relevant
when its future return correlation with the query exceeds a fixed
percentile of the pairwise correlation distribution.  At the 75th
percentile \FASCL{} attains Precision@1 of 0.66 against 0.59 for
Pearson correlation, and the ordering is unchanged at the 50th
percentile (0.78 vs.\ 0.73) and the 90th (0.55 vs.\ 0.48).

\noindent\textbf{Efficiency.}
\FASCL{} embeds the 16,370-sample test set in 4.3\,s with sub-0.01\,ms
per-query search, in the same range as the other compact neural
encoders.  DTW pays 277\,ms per query at search time and the
foundation models 50--390\,s of embedding time, yet both trail \FASCL{} on every
future-behavior metric, so \FASCL{}'s retrieval quality does not come
from additional computation.

\begin{figure*}[t]
  \centering
  \begin{subfigure}{0.46\textwidth}
    \centering
    \includegraphics[width=\linewidth]{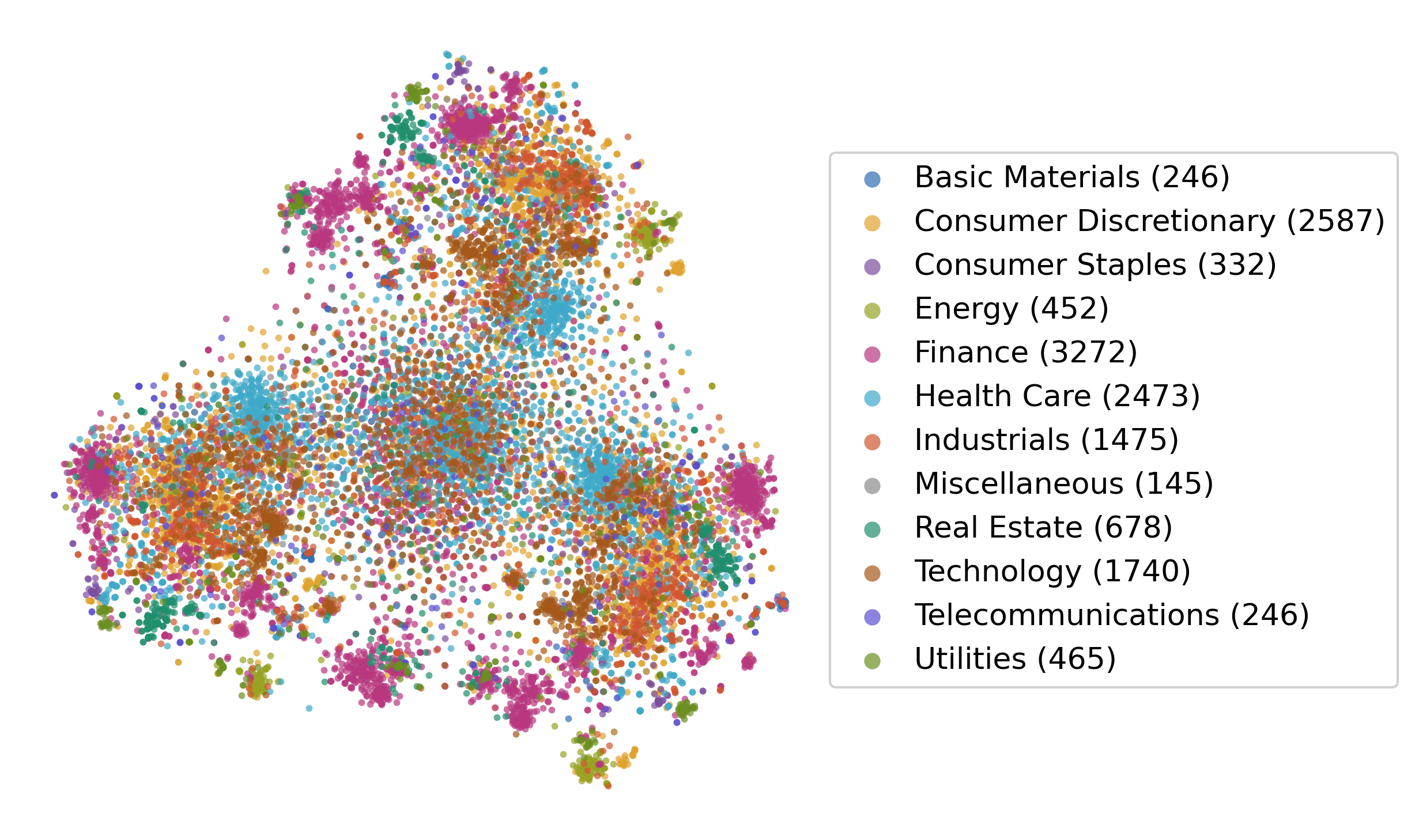}
    \caption{Sector.}
  \end{subfigure}
  \hspace{4pt}
  \begin{subfigure}{0.50\textwidth}
    \centering
    \includegraphics[width=\linewidth]{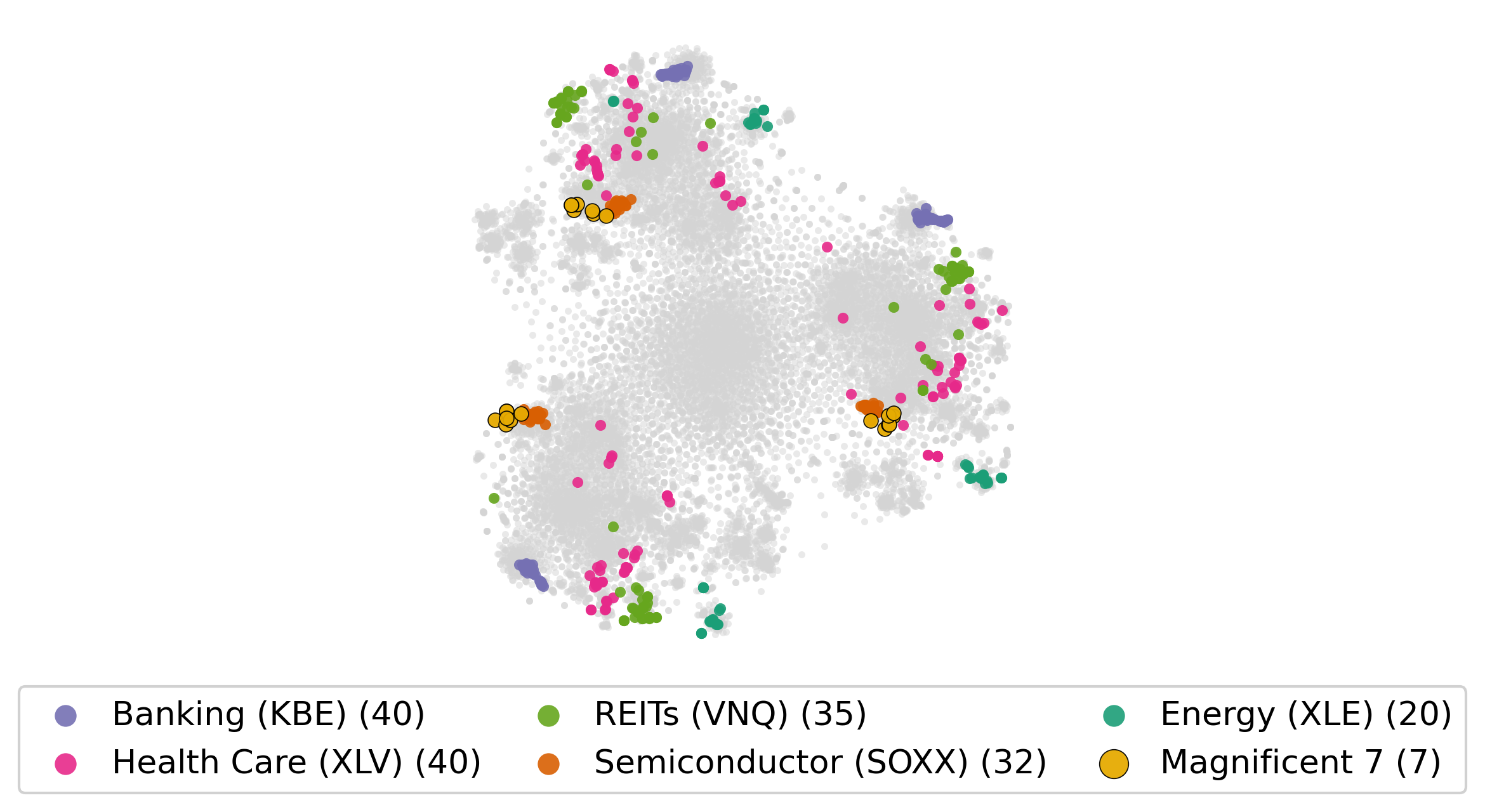}
    \caption{Thematic ETF membership.}
  \end{subfigure}
  \caption{t-SNE projections of \FASCL{} test-set embeddings colored
  (a)~by sector, for the 86\% of samples with sector metadata, and
  (b)~by membership in five thematic ETFs and the Magnificent~7
  (non-members in gray).  No sector, industry, or ETF labels are used
  during training.}
  \vspace{-10pt}
  \label{fig:tsne_cat}
\end{figure*}

\begin{figure}[t]
  \centering
  \includegraphics[width=0.9\linewidth]{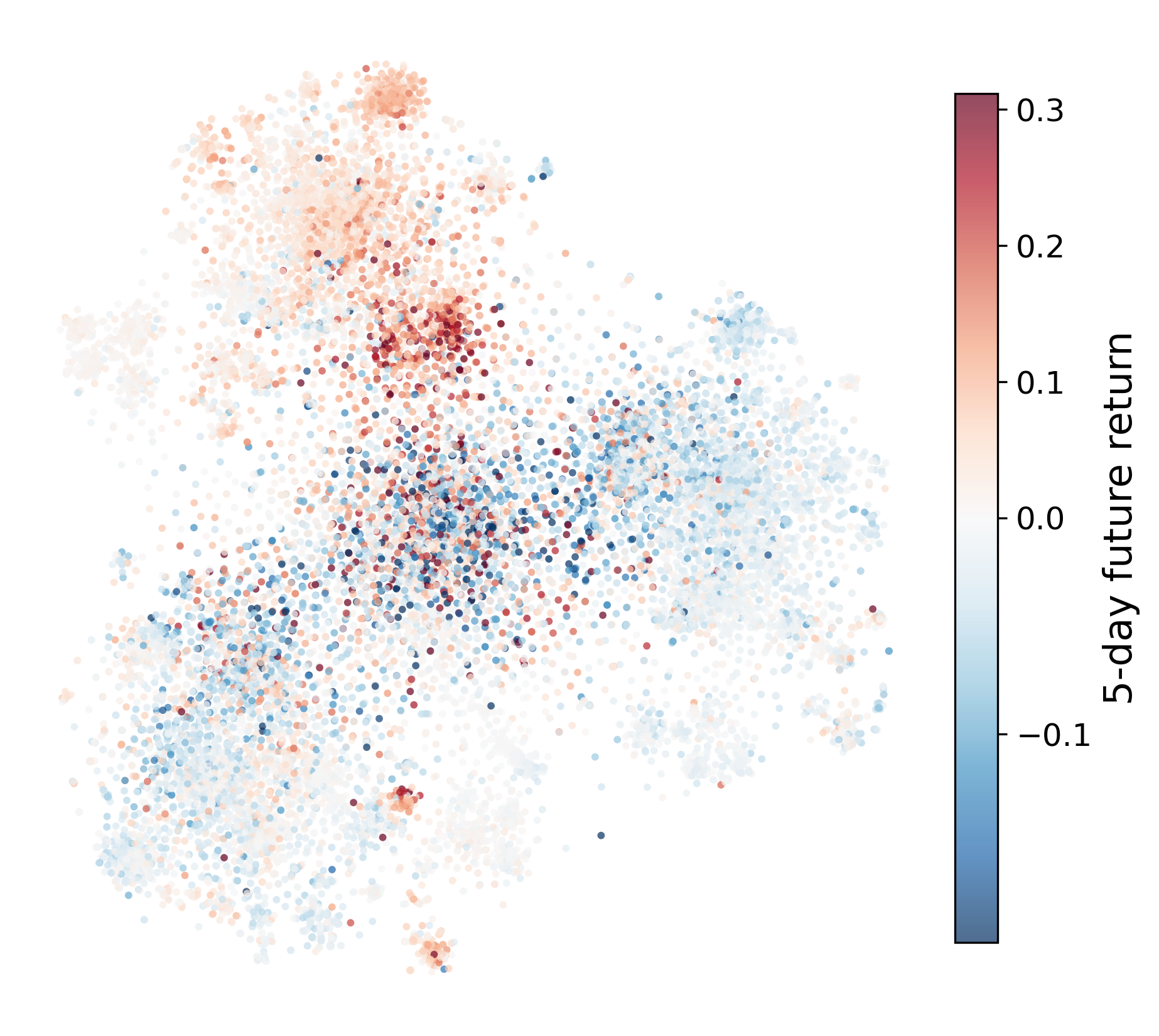}
  \caption{t-SNE projection of \FASCL{} test-set embeddings
  (16{,}370 samples) colored by realized 5-day future return on a
  zero-centered diverging scale.}
  \label{fig:tsne}
\end{figure}

\subsection{Ablation Study}
\label{sec:ablation}

Table~\ref{tab:ablation} isolates the contribution of each design
choice.  \FASCL{} trains with the soft contrastive loss alone.  A
natural extension is to add an auxiliary regression head, a two-layer
MLP on the embedding that predicts each asset's own cumulative return
at 1, 5, 20, and 60 days under a smooth-L1 loss weighted at 0.1
relative to $\mathcal{L}_{\mathrm{SC}}$; this changes FRC@$K$ by less
than 1\% at every~$K$ (second row), so we keep the simpler objective.
The results form a clear hierarchy: what the
embedding space is aligned to matters far more than how the encoder
is built.

\noindent\textbf{Supervision and loss.}
Removing the contrastive objective entirely, by training the same
encoder to regress future returns pointwise, collapses FRC@1 from
0.443 to 0.167, near the random baseline of 0.141.  Retrieval quality
therefore does not fall out of a generic future-prediction objective;
it requires the pairwise alignment that the contrastive formulation
provides.
Keeping the soft contrastive form but deriving targets from
observation-space cosine similarity instead of future return
correlation costs 35\% of FRC@1 (0.288) and drops label-free sector
precision from 53.4\% to 32.6\% at $K{=}1$.  This variant changes
nothing except the supervision source, and it lands inside the
input-space band of Table~\ref{tab:test_frc_sp}, confirming that the
alignment target, not the loss machinery around it, is what separates
\FASCL{} from observation-space methods.
The remaining loss variants have small effects: collapsing the soft
target distribution to binary InfoNCE labels costs 3.8\% at $K{=}1$,
and raising the loss temperature from 0.01 to the conventional 0.07
costs 0.7\%.

\noindent\textbf{Architecture and input.}
Encoder choices are second-order.  Replacing mean pooling with a CLS
token changes FRC@$K$ by at most 4\% relative to the reported
model (slightly positive), so the pooling mechanism is not a critical
choice.  Restricting the input to
close prices alone costs 4.5\%, indicating that the remaining input
channels carry complementary information.  The exception is the
SimCLR-style projection head, which costs 14\% at $K{=}1$: optimizing
the loss on a projected space rather than on the encoder output
degrades the space actually used for retrieval.

\subsection{Robustness}
\label{sec:robustness}

\noindent\textbf{Volatility regimes.}
We split test queries by market volatility during their evaluation
window, using the annualized 20-day realized volatility of the
S\&P~500: a window period whose average exceeds the index's median
over 2024 is labeled high-volatility, which selects one of the three
test periods and 34\% of queries.
\FASCL{} scores FRC@1 of 0.437 there versus 0.445 in the remaining
periods, a decline of under 2\%, and its margin over Pearson
correlation is preserved ($+$9\%, 0.437 vs.\ 0.400).  Retrieval
quality is therefore stable across the volatility conditions present
in the test year, though a single year contains only a limited range
of them.

\noindent\textbf{Temperature sensitivity.}
The loss temperature $\tau{=}0.01$ and target temperature
$\tau_t{=}0.05$ are the only nonstandard hyperparameters in the
objective.  We retrained \FASCL{} at seven grid points
spanning $\tau \in [0.003, 0.1]$ and $\tau_t \in [0.03, 0.2]$ with
all other settings fixed.  FRC@1 moves by at most 1.1\% for
$\tau \in [0.003, 0.05]$ and $\tau_t \in [0.03, 0.1]$, and degrades
gracefully at the extremes (0.428 at $\tau{=}0.1$, 0.419 at
$\tau_t{=}0.2$, versus the reported 0.443), so neither temperature
requires delicate tuning.

\begin{table}[t]
  \centering
  \footnotesize
  \setlength{\tabcolsep}{5pt}
  \begin{tabular}{lcccc}
    \toprule
    Method & K=1 & K=5 & K=10 & K=20 \\
    \midrule
%%GEN:tab_spread BEGIN
    Random & 3.00 & 3.50 & 4.13 & 3.42 \\
    DTW & 4.52 & 3.91 & 3.50 & 3.96 \\
    Pearson Corr. & 5.76 & \second{6.07} & \second{5.73} & \second{5.26} \\
    \midrule
    TS2Vec & 5.49 & 3.43 & 3.90 & 3.63 \\
    CoST & 5.31 & 4.87 & 4.83 & 4.90 \\
    SoftCLT & 4.87 & 4.73 & 4.74 & 4.00 \\
    TNC & 3.08 & 3.10 & 3.69 & 3.30 \\
    \midrule
    PatchTST & 4.36 & 4.55 & 4.47 & 4.42 \\
    TimesNet & 3.02 & 2.96 & 3.33 & 3.54 \\
    iTransformer & 2.93 & 3.93 & 3.45 & 3.84 \\
    \midrule
    SimStock & 4.71 & 4.48 & 4.44 & 4.12 \\
    \midrule
    Chronos & 5.41 & 5.82 & 5.38 & 4.33 \\
    MOMENT & 5.49 & 4.82 & 5.46 & 5.11 \\
    MOMENT (FT) & \second{5.85} & 4.64 & 4.36 & 3.90 \\
    \midrule
    \FASCL{} (Ours) & \best{6.93} & \best{6.18} & \best{6.78} & \best{6.31} \\
%%GEN:tab_spread END
    \midrule
    \multicolumn{5}{l}{\emph{\FASCL{} cost analysis}} \\
%%GEN:tab_spread_costs BEGIN
    Turnover (daily) & 0.32 & 0.32 & 0.32 & 0.31 \\
    Break-even cost (bps) & 11.7 & 8.3 & 9.4 & 9.0 \\
    Net Sharpe @ 5\,bps & 3.98 & 2.49 & 3.18 & 2.84 \\
    Net Sharpe @ 10\,bps & 1.00 & -1.24 & -0.41 & -0.71 \\
%%GEN:tab_spread_costs END
    \bottomrule
  \end{tabular}
  \caption{Spread trading backtest on the test set.  Upper panel:
  annualized gross Sharpe ratio per method; higher values indicate
  tighter co-movement between query and retrieved peers.  Lower panel:
  \FASCL{}'s average daily traded notional per unit of spread exposure,
  the one-way cost at which mean net P\&L reaches zero, and net Sharpe
  under fixed one-way costs charged on traded notional.}
  \vspace{-20pt}
  \label{tab:spread}
\end{table}

\begin{table}[t]
  \centering
  \footnotesize
  \setlength{\tabcolsep}{4pt}
  \begin{tabular}{lcccc}
    \toprule
    Method & K=1 & K=5 & K=10 & K=20 \\
    \midrule
%%GEN:tab_tracking BEGIN
    Random & 0.0384 & 0.0283 & 0.0244 & 0.0224 \\
    DTW & 0.0305 & 0.0243 & 0.0225 & 0.0213 \\
    Pearson Corr. & \second{0.0255} & \second{0.0217} & \second{0.0204} & \second{0.0195} \\
    \midrule
    TS2Vec & 0.0286 & 0.0228 & 0.0216 & 0.0207 \\
    CoST & 0.0311 & 0.0244 & 0.0223 & 0.0209 \\
    SoftCLT & 0.0297 & 0.0240 & 0.0221 & 0.0211 \\
    TNC & 0.0362 & 0.0271 & 0.0241 & 0.0223 \\
    \midrule
    PatchTST & 0.0326 & 0.0259 & 0.0235 & 0.0218 \\
    TimesNet & 0.0331 & 0.0258 & 0.0233 & 0.0217 \\
    iTransformer & 0.0327 & 0.0257 & 0.0231 & 0.0216 \\
    \midrule
    SimStock & 0.0348 & 0.0269 & 0.0239 & 0.0221 \\
    \midrule
    Chronos & 0.0302 & 0.0235 & 0.0218 & 0.0207 \\
    MOMENT & 0.0303 & 0.0241 & 0.0222 & 0.0210 \\
    MOMENT (FT) & 0.0303 & 0.0241 & 0.0222 & 0.0210 \\
    \midrule
    \FASCL{} (Ours) & \best{0.0215} & \best{0.0192} & \best{0.0188} & \best{0.0186} \\
%%GEN:tab_tracking END
    \bottomrule
  \end{tabular}
  \caption{Tracking error (lower is better) of the spread trading
  backtest on the test set.  \FASCL{} achieves the lowest tracking error
  at all basket sizes, with ${\sim}$16\% relative reduction over Pearson
  correlation at $K{=}1$ and ${\sim}$12\% at $K{=}5$.}
  \vspace{-15pt}
  \label{tab:tracking_error}
\end{table}

\subsection{Embedding Visualization}
\label{sec:visualization}
Figures~\ref{fig:tsne_cat} and~\ref{fig:tsne} present t-SNE
projections of \FASCL{} embeddings on the test set, colored by
different attributes to reveal the structure of the learned
representation space.

\noindent\textbf{Emergent sector structure (Figure~\ref{fig:tsne_cat}a).}
Colored by sector, the space contains many tight same-sector clusters
(visible for Finance, Health Care, Energy, Real Estate, and Technology)
although \FASCL{} receives no categorical metadata during training.
This structure emerges because firms sharing common factor exposures,
such as interest-rate sensitivity for financials, tend to exhibit
correlated future returns, which the soft contrastive loss encodes into
embedding proximity.  Sector identity is recovered only where it
predicts co-movement, consistent with the gap between \FASCL{} and the
sector-informed SimStock on FRC@$K$ versus SP@$K$
(Section~\ref{sec:main_results}).

\noindent\textbf{Thematic structure (Figure~\ref{fig:tsne_cat}b).}
Overlaying the constituents of five thematic ETFs and the
Magnificent~7 sharpens the picture: semiconductors, banks, energy, and
REITs each concentrate in a few tight clusters, and the Magnificent~7
sits directly beside the semiconductor group although its members span
three sectors.  Because the same-period protocol only compares windows
within a period, the projection separates the three test periods into
distinct regions; the same thematic clusters recur in each of them.

\noindent\textbf{Future return structure (Figure~\ref{fig:tsne}).}
Colored by realized 5-day future return, the embedding space shows a
broad spatial gradient: samples with positive returns concentrate in
one region and negative-return samples in another, with locally mixed
signs in between.

\subsection{Downstream Application: Spread Trading}
\label{sec:spread}

To evaluate whether future-aligned retrieval translates into practical
investment value, we conduct a stylized spread trading
backtest~\cite{gatev2006pairs} designed to measure \emph{signal quality}
rather than realized profitability.  We report gross Sharpe ratios for
all methods, and for \FASCL{} additionally the strategy's turnover and
its performance net of transaction costs.
For each query asset~$a_q$ at time~$t$, the top-$K$ retrieved peers form
an equal-weight basket.  The daily spread return is defined as:
\begin{equation}
  s_{q,d} = r_{q,d} - \frac{1}{K}\sum_{k=1}^{K} r_{k,d},
  \quad d = 1, \ldots, H,
\end{equation}
where $r_{q,d}$ and $r_{k,d}$ are the daily returns of the query and
the $k$-th peer, respectively.  A mean-reversion signal trades against
the cumulative spread: the position at day~$d{+}1$ is
$-\mathrm{sign}\!\bigl(\sum_{d'=1}^{d} s_{q,d'}\bigr)$,
and the daily P\&L is the product of this position and the next-day
spread.  Portfolio-level P\&L is obtained by averaging daily P\&L
across all queries, and we report the annualized Sharpe ratio
($\text{mean}/\text{std} \times \sqrt{252}$) of the resulting
portfolio time series.  Daily returns entering the backtest are
clipped at $\pm$50\% to guard against data artifacts such as
mis-adjusted splits.

Table~\ref{tab:spread} shows gross Sharpe ratios across basket
sizes~$K$.  \FASCL{} attains the highest value at every~$K$
(6.93/6.18/6.78/6.31), ahead of the strongest alternatives at each~$K$ (fine-tuned MOMENT 5.85 at $K{=}1$; Pearson 6.07 at
$K{=}5$).  The soft contrastive loss aligns the \emph{entire}
similarity ranking with future return correlations, so even
lower-ranked peers co-move meaningfully with the query, which is what
basket-based strategies require.

\noindent\textbf{Turnover and transaction costs.}
The mean-reversion position is the sign of the cumulative spread,
recomputed daily, so it can flip often; each flip trades both legs.
The lower panel of Table~\ref{tab:spread} reports the resulting
turnover and cost sensitivity for \FASCL{}: average daily traded
notional is ${\sim}0.32$ per unit of spread exposure, the strategy
remains profitable up to a break-even one-way cost of 8--12\,bps
depending on~$K$, and at a 5\,bps cost assumption (conservative for
liquid US equities) the net Sharpe stays between 2.49 and 3.98.
The signal therefore survives realistic cost levels, though a stylized
backtest omits market impact and capacity effects.

\noindent\textbf{Tracking error.}
Table~\ref{tab:tracking_error} reports the median across queries of
the standard deviation of daily spread returns, a direct measure of
how tightly retrieved peers co-move with the query.  \FASCL{} achieves the lowest
tracking error at every basket size, 16\% below Pearson correlation at
$K{=}1$ and 12\% at $K{=}5$: its peer baskets not only produce
stronger mean-reversion signals but also exhibit lower residual
volatility.

%% =============================================================================
%% 6. CONCLUSION
%% =============================================================================

% Spread and tracking tables: declared at the section's end so they land
% on the following page instead of stacking with Figure 1.

\section{Conclusion}
\label{sec:conclusion}

We proposed \FASCL{}, a representation learning framework that aligns
embedding similarity with pairwise future return correlations through a
soft contrastive loss, establishing \emph{future-aligned asset retrieval}
as a well-defined research problem.  Against 14 baselines on 5,631
US-listed securities, \FASCL{} attains the best score in every FRC@$K$
and IC@$K$ cell and in 13 of 16 trend consistency cells, and the highest
gross Sharpe at every basket size in a spread trading backtest.  The
ablations locate the source: removing the contrastive objective costs
62\% of FRC@1 and replacing future-aligned targets with
observation-aligned ones costs 35\%, while architectural choices matter
far less.  The evaluation protocol and the baseline suite are further
contributions for research on financial asset retrieval.

\noindent\textbf{Limitations and future work.}
The supervision target defines a design space we have only begun to
explore: any pairwise dependence measure can replace Pearson
correlation without changing the framework, whether rank-based
concordance such as Kendall's~$\tau$, tail dependence for risk
management, or downstream-utility signals such as spread convergence.
Pearson correlation is also scale-invariant and the temperature-scaled
softmax concentrates gradient on highly correlated pairs, so
volatility-aware targets and an explicit treatment of anti-correlated
assets are natural refinements for trading and hedging.
Finally, this work addresses \emph{same-period} retrieval; matching a
current pattern against other periods (\emph{cross-time retrieval}) and
validation on non-US markets remain future work.

%% =============================================================================
%% ETHICAL CONSIDERATIONS
%% =============================================================================
\section*{Ethical Considerations}

This work uses only publicly available market price and volume data
for exchange-listed securities; no personal or user data is involved.
Future-aligned retrieval is intended as decision support, surfacing
behaviorally related assets for diversification, hedging, and risk
analysis.  Its outputs are not investment advice, and the spread
trading backtest is a stylized signal-quality measure that omits
market impact and capacity constraints, so any strategy built on it
would need further risk controls and cost modeling.  As with any
co-movement signal, broad adoption could concentrate trades among
assets identified as related; the method surfaces relationships
already present in market prices rather than generating novel
forecasts, which limits but does not eliminate that risk.

%% =============================================================================
%% REFERENCES
%% =============================================================================
\bibliographystyle{ACM-Reference-Format}
\bibliography{references}

%% =============================================================================
%% SUPPLEMENTARY MATERIAL (appended after the references)
%% =============================================================================
\clearpage
\setcounter{section}{0}
\setcounter{table}{0}
\setcounter{figure}{0}
\renewcommand\thesection{S\arabic{section}}
\renewcommand\thetable{S\arabic{table}}
\renewcommand\thefigure{S\arabic{figure}}

\section*{Supplementary Material}

\noindent This supplement collects extended results and configuration
detail that support the main paper but did not fit its page budget:
the ablation grid scored on two further metrics
(Section~\ref{sec:supp:ablation}), the full temperature sensitivity
grid (Section~\ref{sec:supp:tau}), the training configurations of the
learning-based baselines (Section~\ref{sec:supp:baselines}), the
computational cost of every method (Section~\ref{sec:supp:cost}), the
embedding structure at each return horizon
(Section~\ref{sec:supp:horizons}), and the constituent lists behind the
thematic-ETF visualization panel (Section~\ref{sec:supp:etf}).  All retrieval numbers are produced by
the same evaluation pipeline as the tables in the body.

\section{Extended Ablation: TC and SP}
\label{sec:supp:ablation}

Table~\ref{tab:supp:ablation_ext} scores the ablation variants of Table~\ref{tab:ablation} on trend consistency at the 20-day horizon
(TC@$K$) and sector precision (SP@$K$), using the same checkpoints and
the same evaluator.  The supervision hierarchy reported there for
FRC@$K$ reappears on both metrics: removing the contrastive objective
collapses TC@1 to the 54\% range and SP@1 to near-random (16.6\%),
and switching to observation-aligned targets loses 7.5\,pp of TC@1
and 20.8\,pp of SP@1 relative to the reported model, while the
architecture- and input-side variants stay within about 2.5\,pp of it
on TC@$K$.

\begin{table}[h]
  \centering
  \footnotesize
  \setlength{\tabcolsep}{3pt}
  \begin{tabular}{lcccccccc}
    \toprule
    & \multicolumn{4}{c}{TC@$K$ (20d, \%)} & \multicolumn{4}{c}{SP@$K$ (\%)} \\
    \cmidrule(lr){2-5}\cmidrule(lr){6-9}
    Variant & K=1 & K=5 & K=10 & K=20 & K=1 & K=5 & K=10 & K=20 \\
    \midrule
%%GEN:tab_ablation_ext BEGIN (verbatim from output/wsdm/paper_tables/tab_ablation_ext.tex)
    \FASCL{} & 67.1 & 64.6 & 64.0 & 63.5 & 53.4 & 50.2 & 47.8 & 45.3 \\
    \;+ auxiliary regression head & 67.5 & 65.9 & 65.1 & 64.4 & 53.7 & 50.9 & 48.4 & 45.7 \\
    \midrule
    Observation-aligned targets & 59.6 & 58.4 & 58.1 & 57.4 & 32.6 & 28.7 & 27.1 & 25.7 \\
    Hard InfoNCE targets & 66.7 & 65.3 & 64.5 & 63.5 & 52.4 & 48.5 & 46.1 & 43.6 \\
    Return regression only & 54.6 & 54.0 & 53.8 & 53.8 & 16.6 & 16.5 & 16.2 & 16.0 \\
    Loss temperature $\tau{=}0.07$ & 68.1 & 65.9 & 64.8 & 64.1 & 53.4 & 49.3 & 47.0 & 44.4 \\
    \midrule
    CLS pooling & 65.4 & 62.6 & 62.3 & 62.4 & 53.3 & 49.5 & 47.2 & 44.9 \\
    With projection head & 64.6 & 63.0 & 62.4 & 61.6 & 44.7 & 40.4 & 38.3 & 36.2 \\
    Close price only & 66.5 & 65.5 & 64.6 & 63.8 & 46.8 & 44.3 & 42.0 & 39.7 \\
%%GEN:tab_ablation_ext END
    \bottomrule
  \end{tabular}
  \caption{Ablation variants of Table~\ref{tab:ablation} on test-set
  TC@$K$ at the 20-day horizon and SP@$K$.  \FASCL{} uses the soft
  contrastive loss alone; every row below the first rule keeps the
  combined configuration and modifies exactly one component.  SP@$K$
  is computed over the 86\% of samples with sector metadata.}
  \label{tab:supp:ablation_ext}
\end{table}

\section{Temperature Sensitivity: Full Grid}
\label{sec:supp:tau}

Table~\ref{tab:supp:tau} reports the complete temperature grid
summarized in Section~\ref{sec:robustness}: seven retrained
variants of the reported model with only the loss temperature~$\tau$
and target temperature~$\tau_t$ changed, plus the reported model
itself.  FRC@1 moves by at most 1.1\% for $\tau \in [0.003, 0.05]$
and $\tau_t \in [0.03, 0.1]$, and degrades gracefully at the extremes.

\begin{table}[h]
  \centering
  \footnotesize
  \setlength{\tabcolsep}{5pt}
  \begin{tabular}{cccccc}
    \toprule
    $\tau$ & $\tau_t$ & FRC@1 & FRC@5 & FRC@10 & FRC@20 \\
    \midrule
%%GEN:tab_tau BEGIN (verbatim from output/wsdm/paper_tables/tab_tau.tex)
    0.003 & 0.05 & 0.4413 & 0.4220 & 0.4086 & 0.3912 \\
    0.005 & 0.05 & 0.4461 & 0.4266 & 0.4119 & 0.3939 \\
    0.01 & 0.05$^{\dagger}$ & 0.4427 & 0.4226 & 0.4083 & 0.3896 \\
    0.05 & 0.05 & 0.4414 & 0.4191 & 0.4055 & 0.3893 \\
    0.1 & 0.05 & 0.4284 & 0.4047 & 0.3914 & 0.3766 \\
    0.01 & 0.03 & 0.4416 & 0.4244 & 0.4077 & 0.3864 \\
    0.01 & 0.1 & 0.4379 & 0.4149 & 0.4020 & 0.3870 \\
    0.01 & 0.2 & 0.4186 & 0.3966 & 0.3855 & 0.3731 \\
%%GEN:tab_tau END
    \bottomrule
  \end{tabular}
  \caption{Test-set FRC@$K$ across the temperature grid.  Each row is
  a full retraining on the same data and protocol (100 epochs);
  $^{\dagger}$ marks the reported configuration.}
  \label{tab:supp:tau}
\end{table}

\section{Baseline Training Configurations}
\label{sec:supp:baselines}

All training-based baselines use the same temporal split (2010--2022),
6-channel input, and 64-day windows, with learning rate $10^{-3}$
except the MOMENT fine-tune ($10^{-5}$).
Table~\ref{tab:supp:baselines} lists the per-method configurations.

\begin{table}[h]
  \centering
  \footnotesize
  \setlength{\tabcolsep}{2.5pt}
  \begin{tabular}{llllll}
    \toprule
    Method & Encoder & Loss & Epochs & Batch & Dim \\
    \midrule
    TS2Vec & Dilated CNN (d10, h64) & Hierarchical CL & 200 & 16 & 320 \\
    CoST & Dilated CNN (d10, h64) & MoCo CL & 200 & 16 & 320 \\
    SoftCLT & Dilated CNN (d10, h64) & Soft CL & 200 & 16 & 320 \\
    TNC & GRU (hidden 100) & Neighborhood CL & 100 & 10 & 64 \\
    PatchTST & Transformer (3 layers) & MSE & 10 & 32 & 128 \\
    TimesNet & TimesBlock (2 layers) & MSE & 10 & 32 & 64 \\
    iTransformer & Inverted Tr.\ (2 layers) & MSE & 10 & 32 & 128 \\
    SimStock & HyperAttention + LSTM & Triplet & 3 & 256 & 256 \\
    MOMENT (FT) & T5-large (342M FT) & MSE & 5 & 16 & 1024 \\
    \bottomrule
  \end{tabular}
  \caption{Training configurations of the learning-based baselines
  (d = depth, h = hidden width).}
  \label{tab:supp:baselines}
\end{table}

\noindent Additional detail:
\begin{itemize}
\item SimStock additionally consumes each asset's sector label as a
  categorical input embedding, following its original design; this is
  what places its sector precision near the ceiling in Table~\ref{tab:test_frc_sp}.
\item TS2Vec, CoST, and SoftCLT share the same dilated convolutional
  encoder.  SoftCLT uses the official implementation with
  cosine-distance soft labels; the global distance range for label
  normalization is estimated from 400k sampled pairs.
\item The forecasting models (PatchTST, TimesNet, iTransformer) are
  trained for 16-step-ahead prediction; encoder representations are
  extracted for retrieval with the models' own instance normalization
  applied at embed time.
\item For every objective that regresses raw values (the forecasting
  models and the MOMENT fine-tune), inputs and targets are normalized
  per window with the input window's channel statistics and winsorized
  at $10\sigma$; without this, the volume and trading-value channels
  dominate the MSE by several orders of magnitude.
\item MOMENT and Chronos (zero-shot) use the publicly released
  pre-trained weights (\texttt{AutonLab/MOMENT-1-large},
  \texttt{amazon/chronos-t5-base}) with no training.  The statistical
  baselines (Random, Pearson, DTW) require no training.
\end{itemize}

\section{Embedding and Search Cost}
\label{sec:supp:cost}

Table~\ref{tab:supp:cost} reports the wall-clock cost of the two
retrieval stages for every method: embedding the 16{,}370-sample test
set once, and per-query search over the same-period candidate pool.
Values that round to zero at two decimals are shown as ${<}0.01$; an
em dash marks methods with no embedding stage.

\begin{table}[h]
  \centering
  \footnotesize
  \begin{tabular}{lcc}
    \toprule
    Method & Embed (s, test set) & Search (ms/query) \\
    \midrule
    Random & --- & ${<}0.01$ \\
    DTW & --- & 276.65 \\
    Pearson Corr. & --- & ${<}0.01$ \\
    \midrule
    TS2Vec & 5.49 & ${<}0.01$ \\
    CoST & 8.07 & ${<}0.01$ \\
    SoftCLT & 3.27 & ${<}0.01$ \\
    TNC & 2.86 & ${<}0.01$ \\
    \midrule
    PatchTST & 3.54 & ${<}0.01$ \\
    TimesNet & 5.65 & ${<}0.01$ \\
    iTransformer & 3.46 & ${<}0.01$ \\
    \midrule
    SimStock & 72.92 & ${<}0.01$ \\
    \midrule
    Chronos & 118.75 & ${<}0.01$ \\
    MOMENT & 49.72 & ${<}0.01$ \\
    MOMENT (FT) & 390.24 & ${<}0.01$ \\
    \midrule
    \FASCL{} & 4.34 & ${<}0.01$ \\
    \bottomrule
  \end{tabular}
  \caption{Embedding and search cost per method.}
  \label{tab:supp:cost}
\end{table}

\section{Future-Return Structure Across Horizons}
\label{sec:supp:horizons}

\begin{figure*}[t]
  \centering
  \begin{subfigure}{0.42\textwidth}
    \includegraphics[width=\linewidth]{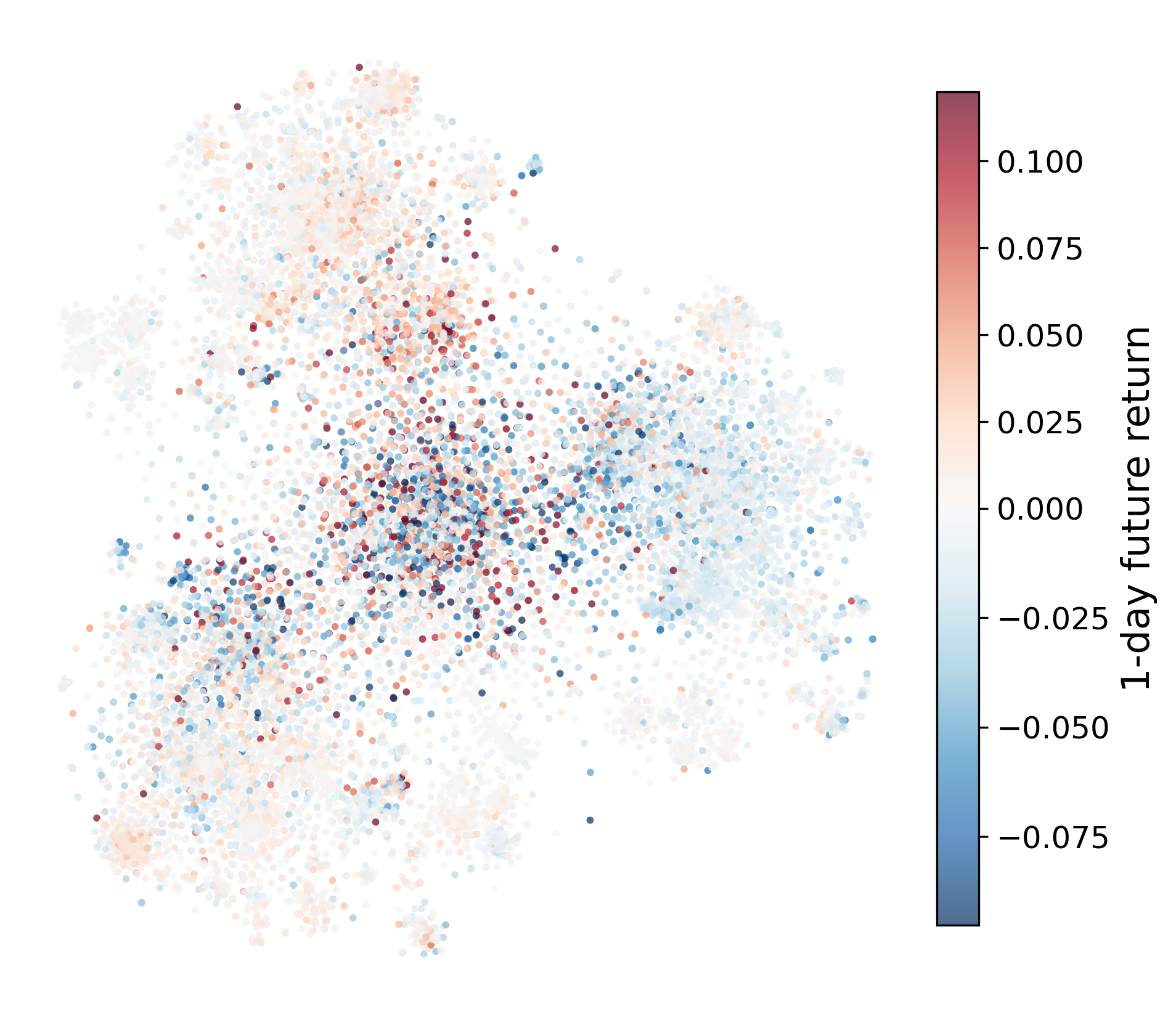}
    \caption{1-day horizon.}
  \end{subfigure}\hfill
  \begin{subfigure}{0.42\textwidth}
    \includegraphics[width=\linewidth]{figs/tsne/b1_return_5d.png}
    \caption{5-day horizon (Figure~\ref{fig:tsne}).}
  \end{subfigure}\\[4pt]
  \begin{subfigure}{0.42\textwidth}
    \includegraphics[width=\linewidth]{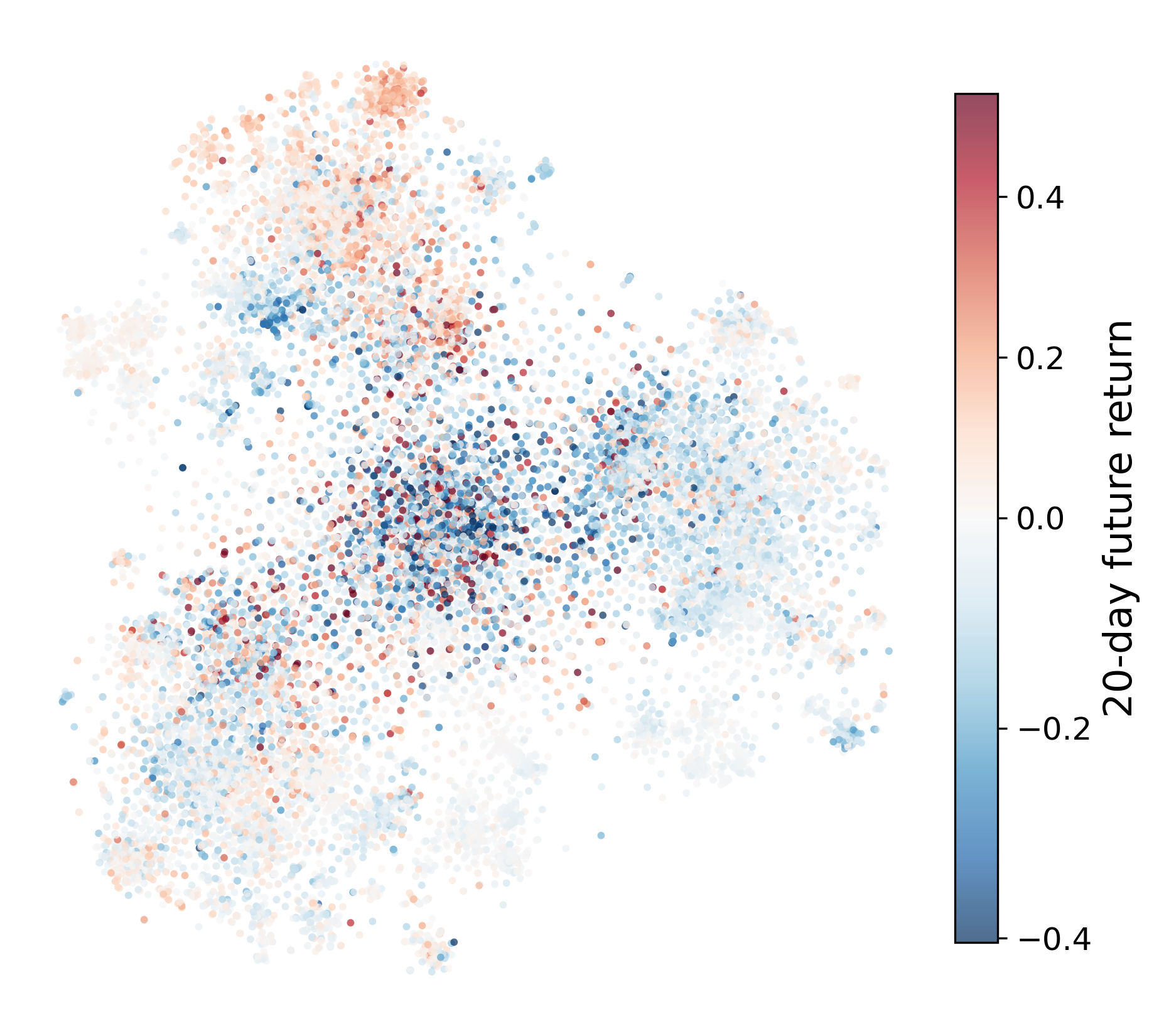}
    \caption{20-day horizon.}
  \end{subfigure}\hfill
  \begin{subfigure}{0.42\textwidth}
    \includegraphics[width=\linewidth]{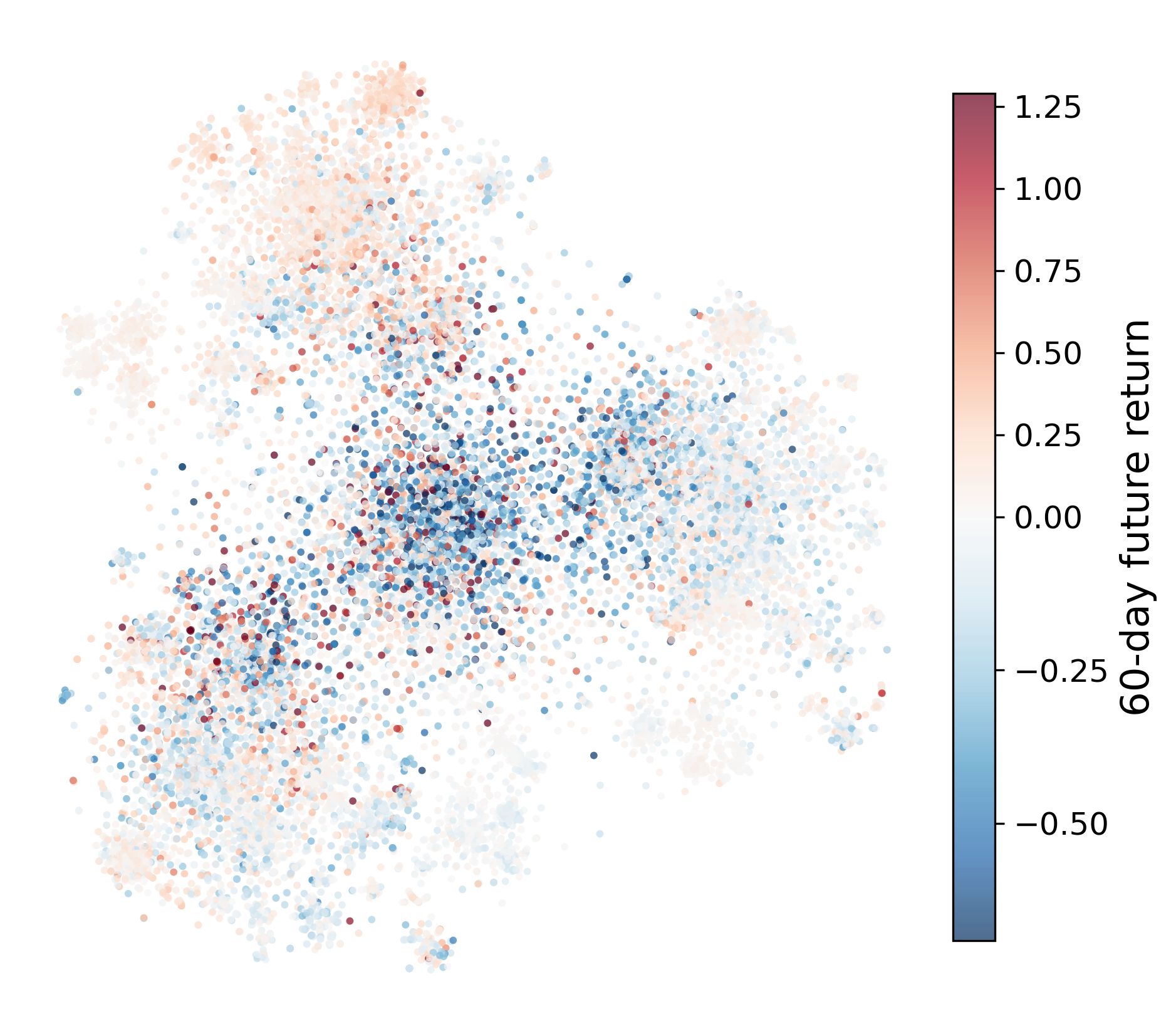}
    \caption{60-day horizon.}
  \end{subfigure}
  \caption{One t-SNE projection of \FASCL{} test-set embeddings
  (16{,}370 samples), colored by realized future return at four
  horizons on zero-centered diverging scales with 1/99-percentile
  clipping.  All four panels share the same projection and the same
  reported checkpoint.}
  \label{fig:supp:horizons}
\end{figure*}

Figure~\ref{fig:supp:horizons} colors a single t-SNE projection of the
test-set embeddings by realized future return at all four horizons;
the 5-day panel is identical to Figure~\ref{fig:tsne}.  The broad
arrangement persists across horizons --- positive returns concentrated
in the upper region, negative mass in the center and right --- and its
visual strength follows the horizons' signal content: clearest at 5
and 20 days, fainter at 60 days, and faintest at 1 day, where
single-day returns are mostly noise.  This ordering matches the
multi-horizon retrieval results in Tables~\ref{tab:test_tc} and~\ref{tab:test_ic}
(TC@1 of 69.7, 67.1, 66.0, and 60.4 at the 5-, 20-, 60-, and 1-day
horizons).

\section{Thematic-ETF Constituent Lists}
\label{sec:supp:etf}

The lists below are the ticker sets behind the ETF-membership panel of
the embedding visualization in Figure~\ref{fig:tsne_cat}(b).  They are
snapshots of the funds' published holdings (top holdings for the
cap-weighted funds, a broad slice for the equal-weighted KBE), not
point-in-time reconstructions: each list is intersected with the
5{,}576-ticker test universe at plot time, so constituents outside the
universe --- for example companies acquired or delisted before the data
snapshot --- drop out, and the panel legend reports the per-group
counts after this intersection.  The exact lists are hardcoded in
\texttt{scripts/paper\_tables/gen\_tsne\_etf.py} in the released code.

\medskip
\noindent\textbf{Semiconductor (SOXX):} NVDA AVGO AMD QCOM TXN INTC MU
AMAT LRCX KLAC MRVL ADI NXPI MCHP ON MPWR TER SWKS QRVO ENTG LSCC COHR
AMKR RMBS ONTO ASML TSM UMC ASX STM CRUS POWI FORM

\medskip
\noindent\textbf{Energy (XLE):} XOM CVX COP EOG SLB MPC PSX VLO WMB OKE
KMI HES OXY HAL DVN BKR FANG TRGP CTRA MRO APA EQT

\medskip
\noindent\textbf{Banking (KBE):} JPM BAC WFC C USB PNC TFC MTB FITB
HBAN RF KEY CFG ZION CMA WAL EWBC FHN SNV PB CBSH WBS PNFP GBCI ONB
UMBF FNB HOMB SSB VLY COLB UBSI WTFC CADE ABCB AUB BOKF CFR TCBI FULT
HWC OZK BANR CATY INDB PPBI

\medskip
\noindent\textbf{Health Care (XLV):} LLY UNH JNJ ABBV MRK TMO ABT DHR
PFE AMGN ISRG ELV VRTX SYK BSX REGN GILD CI ZTS CVS BDX MCK HCA MRNA
BIIB IDXX IQV A EW DXCM HUM MDT CNC BAX WAT RMD STE COO COR VTRS

\medskip
\noindent\textbf{REITs (VNQ):} PLD AMT EQIX WELL CCI PSA SPG O DLR VICI
EXR AVB CBRE IRM EQR WY SBAC INVH VTR ESS MAA ARE KIM DOC UDR HST REG
BXP CPT FRT AMH ELS LAMR CUBE REXR

\medskip
\noindent\textbf{Magnificent 7:} AAPL MSFT GOOGL GOOG AMZN NVDA META
TSLA

\end{document}